\documentstyle[aps,preprint,epsf]{revtex}
\tightenlines

\def\bild#1#2{    
        \vspace*{-5mm}
        \begin{center}
        \begin{math}
        \epsfxsize#2cm
        \epsffile{#1}
        \end{math}
        \end{center}
        }

\begin{document}
\draft
\title{SU(3) Chiral Dynamics with Coupled Channels:\\ Eta and Kaon
Photoproduction}
\author{N. Kaiser, T. Waas and W. Weise}
\address{Physik Department, Technische Universit\"{a}t M\"{u}nchen\\
   Institut f\"{u}r Theoretische Physik, D-85747 Garching, Germany}

\bigskip

\bigskip

\maketitle
\begin{abstract}
We identify the leading s-wave amplitudes of the SU(3) chiral meson-baryon 
Lagrangian with an effective coupled-channel potential which is iterated in a 
Lippmann-Schwinger equation. The strangeness $S=-1$ resonance $\Lambda(1405)$
and the $S_{11}(1535)$ nucleon resonance emerge as quasi-bound states of
anti-kaon/nucleon and  kaon/$\Sigma$-hyperon. Our approach to meson
photoproduction introduces no new parameters. By adjusting a few finite range
parameters we are able to simultaneously describe a large amount of low energy
data. These  include the cross sections of $K^-p$ elastic and inelastic
scattering, the cross sections of eta meson and kaon photoproduction from
nucleons as well as those of pion  induced production of etas and kaons (16
different reaction channels altogether).
\end{abstract}

\vspace{0in}
\vfill
\noindent $^*${\it Work supported in part by BMBF and GSI}

\newpage

\section{Introduction}
Over the last few years there has been renewed interest in the
photoproduction of eta mesons and kaons from nucleons. At MAMI (Mainz) very
precise differential cross sections  for the reaction $\gamma p \to \eta p$
have been measured from threshold at 707 MeV up to 800 MeV  photon lab energy
\cite{krusche}. The nearly isotropic angular distributions show a clear
dominance of the s-wave amplitude (electric dipole) in this energy range. At 
ELSA (Bonn) an analogous $\eta$-electroproduction experiment has been performed
\cite{schoch} at higher beam energies but with very low virtual photon momentum
transfer, $q^2=-0.056$ GeV$^2$, thus the combined data cover the whole energy 
range of the nucleon resonance $S_{11}(1535)$. The latter has the outstanding
feature of a strong $\eta N$ decay \cite{pdgroup} which is made responsible for
the observed large cross sections. Recently the incoherent  
$\eta$-photoproduction from the deuteron has also been measured at MAMI
\cite{krusched} which allows for a preliminary extraction of the  $\gamma n \to
\eta n$ cross sections \cite{kruschepriv}. Upcoming  coincidence measurements 
of the $\eta$-meson together with a recoiling nucleon will reduce the present 
uncertainties coming from the deuteron structure. At ELSA there is an ongoing  
program to measure strangeness production with photons from proton targets.
Cross sections for the reactions $\gamma p \to K^+\Lambda$
and  $\gamma p \to K^+ \Sigma^0$ have been measured with improved accuracy from
the respective thresholds at 911 and 1046 MeV photon lab energy up to 1.5 GeV
together with angular distributions and recoil hyperon polarizations
\cite{bockhorst}. The analysis of the neutral kaon channel $\gamma p \to K^0 
\Sigma^+$ (considering the same energy range and observables) is presently 
performed \cite{goers} and will lead to a substantial improvement of the data 
base. The ultimate aim of these experimental investigations is a complete
multipole analysis in the low energy region and in particular a determination
of the s-wave multipole, $E_{0+}$, (i.e. the electric dipole amplitude) close
to threshold. The knowledge of these multipoles will permit crucial tests of 
models for strangeness production. 

Most theoretical models used to describe the abovementioned reactions are based
on an effective Lagrangian approach including Born terms and various (meson and
baryon) resonance exchanges \cite{kohno,adelsack,tiator,mart,sauermann,sauerdr}
with the  coupling constants  partly fixed by independent electromagnetic and 
hadronic data. In the work of \cite{tiator} it was furthermore tried to extract
the $\eta NN$ coupling constant from a best fit to the data and to decide 
whether the $\eta NN$ vertex is of pseudoscalar or pseudovector nature. 
Ref.\cite{sauerdr} used a K-matrix model with parameters adjusted to $S_{11}$ 
partial wave (orbital angular momentum $l=0$, total isospin  $I=1/2$) of 
$\pi N$ scattering and predicted quite successfully the cross sections for pion
induced $\eta$-production $\pi^- p \to \eta n$. The  
$\eta$-photoproduction involves as new parameters the photoexcitation strengths
of two $S_{11}$-resonances which are furthermore constrained by pion
photoproduction in the considered energy range. Ref.\cite{sauerdr} finds a good
description of the MAMI data whereas the ELSA data above the resonance peak are
somewhat underestimated. Whereas resonance models work well for
$\eta$-production the situation is more difficult for kaon production where
several different kaon-hyperon final states are possible. As shown in
\cite{mart} resonance models lead to a notorious overprediction of the $\gamma
p \to K^0 \Sigma^+$ and $\gamma n \to K^+ \Sigma^-$ cross section. Only a
drastic reduction of the $K\Sigma N$ coupling constant to nearly a tenth of
its SU(3) value gives a reasonable fit to all available data. This is clearly 
not a convincing solution to the problem. 

We will use here quite a different approach to eta and kaon photoproduction
(and the related pion induced reactions) not introducing any explicit
resonance. Our starting point is the SU(3) chiral effective meson-baryon
Lagrangian at next-to-leading order, the low energy effective field theory
which respects the symmetries of QCD (in particular chiral symmetry). The 
explicit degrees of freedom are only the baryon and pseudoscalar meson octet
with interactions controlled by chiral symmetry and a low energy expansion.  
As shown in previous work \cite{ksw1,ksw2} the effective Lagrangian predicts a
strong attraction in certain channels such as the $\overline K N$ isospin
$I=0$ and the $K\Sigma$ isospin $I=1/2$ s-waves. If this attraction is
iterated to infinite orders in a potential approach (not performing the
systematic loop expansion of chiral perturbation theory) one can dynamically
generate the $\Lambda(1405)$ and the $S_{11}(1535)$ as quasi-bound
meson-baryon states with all properties attributed to these resonances. The
purpose of this paper is to extend the coupled channel potential approach to
meson photo and electroproduction. To the order we are working this extension 
does not introduce any further parameter compared to the pure strong
interaction case. It is then quite non-trivial to find a good description of 
so many available photon and pion induced data for this multi-channel problem 
with just a few free parameters. For both the strong meson-baryon scattering 
and the meson photoproduction processes we will consider only s-waves in this
work. Therefore the comparison with data is necessarily restricted to 
the near threshold region. The s-wave approximation excludes the calculation of
observables like recoil polarization which arises from s- and p-wave
interference terms. The systematic inclusion of p-waves goes beyond the scope 
of this paper and will be considered in the future.

The paper is organized as follows. In the second section we describe the
effective SU(3) chiral meson-baryon Lagrangian at next-to-leading order and we
present  the potential model to calculate strong meson-baryon
scattering and meson photoproduction simultaneously. In the third section we 
discuss our results, the low energy cross sections for the six 
channels present in $K^-$-proton scattering, $K^-p\to K^-p,\, \overline{K^0}n,
\, \pi^0 \Lambda,\, \pi^+\Sigma^-,\, \pi^0 \Sigma^0,\, \pi^- \Sigma^+$, the
cross sections of eta and kaon photoproduction from
protons (and neutrons) $\gamma p \to \eta p,\\ K^+\Lambda,\, K^+\Sigma^0,\, K^0
\Sigma^+$ and $\gamma n \to \eta n$, as well as those of the corresponding pion
induced reactions $\pi^- p \to \eta n,\, K^0 \Lambda,\, K^0 \Sigma^0 , \, K^+
\Sigma^-$ and $\pi^+ p \to K^+ \Sigma^+$. 
We furthermore make a prediction for the longitudinal to transverse ratio in
$\eta$-electroproduction and discuss the nature of the $S_{11}(1535)$-resonance
in our approach.  In the appendix we collect some lengthy formulae.  

\section{Formalism}
\subsection{Effective Chiral Lagrangian}

The tool to investigate the dynamical implications of spontaneous and explicit
chiral symmetry breaking in QCD is the effective chiral Lagrangian. It provides
a non-linear realization of the chiral symmetry group $SU(3)_L\times SU(3)_R$
in terms of the effective low energy degrees of freedom, which are the
pseudoscalar Goldstone bosons ($\pi,\,K,\,\eta$) and the octet baryons ($N,\,
\Lambda,\,\Sigma,\,\Xi$). The effective Lagrangian can be written  generally as
\cite{revs} 
\begin{equation}
  {\cal L} = {\cal L}^{(1)}_{\phi B} + {\cal L}^{(2)}_{\phi B} + \cdots  
\end{equation}
corresponding to an expansion in increasing number of derivatives (external
momenta) and quark masses. In the relativistic formalism the leading order 
term reads
\begin{equation}
{\cal L}^{(1)}_{\phi B} = {\rm tr}(\overline B(i\gamma_\mu D^\mu-M_0)B)+F\, 
{\rm tr} (\overline B\gamma_\mu\gamma_5[u^\mu,B])+ D\, {\rm tr}(\overline B
\gamma_\mu\gamma_5 \{u^\mu, B\}) \end{equation}
where
\begin{equation} D^\mu B = \partial^\mu B -ie [Q,B] A^\mu + {1\over 8 f^2}
[[\phi , \partial^\mu \phi],B] + \dots \end{equation}
is the chiral covariant derivative and 
\begin{equation} u^\mu = -{1\over 2f} \partial^\mu \phi + {i e \over 2f} [Q,
\phi]  A^\mu + \dots \end{equation}
is  an axial vector quantity. The $SU(3)$ matrices $\phi$ and $B$ collect the
octet pseudoscalar meson fields and the octet baryon fields, respectively. For 
later use the photon field $A^\mu$ has been included via minimal substitution
with  $Q = {1\over 3} {\rm diag}(2,-1,-1)$ the quark charge operator. The scale
parameter $f$ is the pseudoscalar meson decay constant (in the chiral limit)
which we identify throughout with the pion decay constant $f=92.4$ MeV. 
$F\simeq 0.5$ and $D\simeq 0.8$ are the SU(3) axial vector coupling constants 
subject to the constraint $D+F=g_A=1.26$. The mass $M_0$ is the common octet
baryon mass in the chiral limit, which we identify with an average
octet mass.   

At next-to-leading order the terms relevant for s-wave scattering are 
\begin{eqnarray} {\cal L}_{\phi B}^{(2)} &=& b_D\,{\rm tr}(\overline B\{\chi_+,
B\} ) + b_F\, {\rm tr} (\overline B[\chi_+,B]) + b_0\, {\rm tr}\,(\overline B
B)\,{\rm tr} (\chi_+)  \nonumber \\ &+& 2d_D\, {\rm tr} (\overline B \{ (v\cdot
u)^2, B\} ) + 2d_F\, {\rm tr} (\overline B [(v\cdot u)^2,B]) \nonumber\\ &+& 
2d_0 \, {\rm tr}(\overline BB)\,{\rm tr} ((v \cdot u)^2) + 2d_1\, {\rm tr}
(\overline B v\cdot u) \, {\rm tr}(v \cdot u B) \end{eqnarray}
with 
\begin{equation} \chi_+=2\chi_0-{1\over 4 f^2}\{\phi,\{\phi,\chi_0\}\}+\dots,
\quad \chi_0 = {\rm diag}(m_\pi^2, m_\pi^2, 2m_K^2-m_\pi^2)\,\,.\end{equation} 
The first three terms in eq.(5) are chiral symmetry breaking terms linear in
the quark masses.  Using the Gell-Mann-Oakes-Renner relation for the Goldstone
boson masses these can be expressed through  $m_\pi^2$ and $m_K^2$ as done in
eq.(6). Two of the three parameters $b_D, b_F,b_0$ can be fixed from the mass
splittings in the baryon octet
\begin{eqnarray} M_\Sigma - M_\Lambda &=&  {16\over 3} b_D(m_K^2-m_\pi^2)\,, \quad
M_\Xi - M_N = 8 b_F(m_\pi^2-m_K^2)\,, \nonumber \\ M_\Sigma-M_N &=&
4(b_D-b_F)(m_K^2-m_\pi^2) \,.\end{eqnarray}
In a best fit to the isospin averaged baryon masses using the charged meson 
masses one finds the values $b_D=+0.066$ GeV$^{-1}$ and $b_F=-0.213$
GeV$^{-1}$. The $b_0$-term shifts the whole baryon octet by the same amount, so
one needs a further piece of information to fix $b_0$,  which is  the
pion-nucleon sigma term (empirical value $45\pm 8$ MeV \cite{gls})  
\begin{equation} \sigma_{\pi N} = \langle N|\hat m(\bar u u +\bar dd)|N\rangle
= - 2m_\pi^2 (b_D+b_F+2b_0 )\end{equation}
with $\hat m =(m_u+m_d)/2$ the average light quark mass.
At the same time the strangeness content of the proton is given by
\begin{equation}y=  {2\langle p|\bar s s|p \rangle \over \langle p|\bar uu+\bar
dd|p \rangle } = {2(b_0+b_D-b_F)\over 2b_0+b_D+b_F} \end{equation}
whose "empirical" value is presently $y = 0.2 \pm 0.2$ \cite{gls}. If one stays
to linear order in the quark masses, as done here, then both pieces of
information ($\sigma_{\pi N}$ and $y$) can not be explained by a single value
of $b_0$. We will later actually fit $b_0$ to many scattering data within the
bounds, $-0.52 $ GeV$^{-1}$ $<b_0< -0.28$ GeV$^{-1}$ set by the empirical
$\sigma_{\pi N}$ and $y$. The experimentally unknown kaon-proton sigma term
\begin{equation} \sigma_{Kp} = {1\over 2} (\hat m +m_s) \langle p|\bar uu +\bar
ss|p \rangle =-4 m_K^2 (b_D+b_0) \end{equation} 
can then be estimated to linear order in the quark mass.

The last two lines in eq.(5) comprise the general set of order $q^2$ terms
contributing to s-wave meson-baryon scattering. They are written in the 
heavy baryon language with $v^\mu$ a four-velocity which allows to select a
frame of reference (in our case the meson-baryon center of mass frame). Note
that in comparison to previous work \cite{ksw1,ksw2} we use here the minimal 
set of linearly independent terms. The additional term $d_2\,{\rm tr}(\overline
B v\cdot u B v\cdot u)$ can be expressed through the ones given in eq.(5) using
some trace identities of $SU(3)$. Of course the physical content remains the
same if one works with an overcomplete basis as done in \cite{ksw1,ksw2}. The
parameters $d_D,d_F,d_0,d_1$ are not known a priori, but instead of fitting all
of them from data we put two constraints on them,       
\begin{equation} 4\pi \bigg(1 + {m_\pi \over M_N}\biggr) a^+_{\pi N} = {m_\pi^2
\over f^2 } \biggl(d_D+d_F+2d_0 -4b_0 -2b_F-2 b_D -{g_A^2 \over 4 M_N}
\biggr) + {3g_A^2 m_\pi^3 \over 64 \pi f^4 } \end{equation}
and
\begin{equation} 4\pi \biggl(1 + {m_K \over M_N}\biggr)a^0_{K N} = {m_K^2 \over
f^2 } \biggl(4b_F-4b_0-2d_F+2d_0-d_1 +{D \over M_N}\bigl(F-{D\over 3}\bigr)
\biggr)\,\,. \end{equation}
Here $a^+_{\pi N}$ is the isospin-even $\pi N$ s-wave scattering length and
$a^0_{KN}$ the isospin zero kaon-nucleon s-wave scattering length which are
both very small ($a^+_{\pi N} = (-0.012 \pm 0.06)$ fm \cite{koch}, $a^0_{KN} =
-0.1 \pm 0.1$ fm \cite{martin}). The expression for $a^+_{\pi N}$ includes 
the non-analytic loop correction proportional to $m_\pi^3$ calculated in
\cite{bkmpn}, and we have corrected  sign misprints in the formula for
$a^0_{KN}$ occuring in \cite{ksw1}. In essence the relations eqs.(11,12) imply
that these linear combinations of $b$- and $d$-parameters are an order of
magnitude smaller than the individual entries. This completes the description
of the $SU(3)$ chiral meson-baryon Lagrangian at next-to-leading order and we 
conclude that there are only two combinations of $d$-parameters left
free. These will be fixed in a fit to many scattering data. 

\subsection{Coupled Channel Approach}

Whereas the systematic approach to chiral dynamics is chiral perturbation
theory, a renormalized perturbative loop-expansion, its range of applicability
can be very small in cases where strong resonances lie closely above (or even
slightly below) the reaction threshold. Prominent examples for this are the
isospin $I=0$, strangeness $S=-1$ resonance $\Lambda(1405)$ in $K^-$-proton
scattering, or the $S_{11}(1535)$ nucleon resonance which has an outstandingly
large coupling to the $\eta N$-channel and therefore is an essential ingredient
in the description of $\eta$-photoproduction. In previous work \cite{ksw1,ksw2}
we have shown that the chiral effective Lagrangian is a good
starting point to dynamically generate such resonances. The chiral Lagrangian
predicts strongly attractive forces in the $\overline K N$ isospin 0 and $K
\Sigma$ isospin 1/2 channels. If this strong attraction is iterated
to all orders, e.g. via a Lippmann-Schwinger equation in momentum space or a
local coordinate-space potential description, quasi-bound meson-baryon states
emerge which indeed have all the characteristic properties of the
$\Lambda(1405)$ or the $S_{11}(1535)$ (e.g. the $K\Sigma$ isospin 1/2 
quasi-bound state has a large branching ratio for decaying into $\eta N$).  
The price to be paid in this approach are some additional finite range
parameters, which must be fitted to data. However, since we are dealing with a
multi-channel problem, it is quite non-trivial to find a satisfactory
description of the data in all reaction channels with so few free parameters.

Let us now describe the potential approach to meson-baryon scattering developed
in \cite{ksw1,ksw2} and show how it can be generalized to meson
photoproduction. The indices $i$ and $j$ label the meson-baryon channels 
involved. They are coupled through a potential in momentum space
\begin{equation}
V_{ij} = {\sqrt{M_i M_j}\over 4 \pi f^2 \sqrt{s}}\, C_{ij}\,\,, \end{equation}
where the relative coupling strengths $C_{ij}$ are, up to a factor $-f^{-2}$, 
the corresponding s-wave amplitudes calculated from the $SU(3)$ chiral
meson-baryon Lagrangian eqs.(2,5) to order $q^2$, which means at most quadratic
in the  meson center of mass energy      
\begin{equation} E_i = {s-M_i^2 + m_i^2 \over 2 \sqrt{s}} \end{equation} and
the meson mass. Here $\sqrt s$ is the total center of mass energy and $M_i$ and
$m_i$ stand for the masses of the baryon and meson in channel $i$,
respectively. The potential $V_{ij}$ is iterated to all orders in a
Lippmann-Schwinger equation of the form 
\begin{equation}
T_{ij} = V_{ij} + \sum_n {2\over \pi} \int_0^\infty dl {l^2 \over k_n^2 +i0
 -l^2 } \biggl( {\alpha_n^2 + k_n^2 \over \alpha_n^2 + l^2} \biggr)^2
V_{in} T_{nj}\,\,, \end{equation}
with $T_{ij}$ the resulting $T$-matrix connecting the in- and outgoing channels
$j$ and $i$. In eq.(15) the index $n$ labels the intermediate meson-baryon
states to be summed over and $\vec l$ is the relative momentum of the off-shell
meson-baryon pair in intermediate channel $n$. The propagator used in eq.(15)
is proportional to a (simple) non-relativistic energy denominator with
$k_n=\sqrt{E_n^2-m_n^2}$ the on-shell relative momentum. The potentials derived
from the chiral Lagrangian have zero range since they stem from a contact
interaction. To make the $dl$-integration convergent a form factor
parametrizing  finite range aspects of the potential has to be introduced. This
is done via a dipole-like off-shell form factor $[(\alpha_n^2+k_n^2)/
(\alpha_n^2+l^2)]^2$ in eq.(15) with $\alpha_n$ a finite range parameter for 
each channel $n$. The form chosen here has the property that on-shell, i.e. 
for $l=k_n$, it becomes identical to one. From physical  considerations one
expects the cut-offs $\alpha_n$ to lie in the range $0.3$ GeV  to $1$  GeV 
reminiscent of the scales related to two-pion exchange or vector meson
exchange. We will actually fix the cut-offs $\alpha_n$ in a fit to many
data keeping in mind physically reasonable ranges. We note that other than
dipole form of the off-shell form factor in eq.(15) have led to similarly good
results. The Lippmann-Schwinger equation for the multi-channel
$T$-matrix $T_{ij}$ can be solved in closed form by simple matrix inversion
\begin{equation} T = (1 - V\cdot G)^{-1}\cdot V \,\,, \end{equation}
where $G$ is the diagonal matrix with entries
\begin{equation}
G_n = {k_n^2\over 2 \alpha_n} - {\alpha_n\over 2}-i \,k_n\,\,, \end{equation}
with $k_n=\sqrt{E_n^2-m_n^2}$ and the appropriate analytic continuation
($i|k_n|$ below threshold $E_n<m_n$). The resulting $S$-matrix 
\begin{equation}
S_{ij} =\delta_{ij} -2i  \sqrt{k_i k_j}\, T_{ij} \end{equation}
is exactly unitary in the subspace of the (kinematically) open channels (but
not crossing symmetric) and
the total (s-wave) cross section for the reaction $(j\to i)$ is calculated via
\begin{equation}\sigma_{ij} = 4\pi{k_i\over k_j} |T_{ij}|^2\,\,. \end{equation}
We note that the kinematical prefactor in eq.(13) has been chosen such that in
Born approximation, i.e. $T_{ij}=V_{ij}$, the cross section $\sigma_{ij}$ has 
the proper relativistic flux factor. Furthermore, one can see that the
imaginary part of the Born series eq.(16) truncated at quadratic order in the
potential matrix $V$ agrees with the one of a one-loop calculation in chiral
perturbation theory. This is so because $M_n k_n/4\pi\sqrt s$ is the 
invariant two-particle phase space and the chosen off-shell form factor is
unity on-shell. However, the real parts do not show chiral logarithms which
would result from a proper evaluation of four-dimensional loop integrals. 

This concludes the general description of our coupled channel approach. We will
first apply it to the six channel problem of $K^-p$ scattering (involving the
channels $\pi^+\Sigma^-,\, \pi^0 \Sigma^0,\, \pi^- \Sigma^+,\, \pi^0 \Lambda,\,
K^-p,\,\overline{K^0}n$). The corresponding potential strengths $C_{ij}$ can
be found in appendix B of \cite{ksw1}, setting $d_2=0$. Secondly we use it for
the four-channel system of $\pi N,\,\eta N,\,K\Lambda,\,K \Sigma$ states with
total isospin 1/2 and the two channel system of $\pi N,\, K\Sigma$ states with
total isospin 3/2, with the corresponding $C_{ij}$ given in the appendix. 

\subsection{Meson Photo- and Electroproduction}

We now extend the same formalism to s-wave meson photoproduction. As in
\cite{sieg} our basic assumption is  that the s-wave photoproduction process 
can by described by a Lippmann-Schwinger equation. In complete analogy
to our description of the strong interaction we will identify the s-wave
photoproduction potential (named $B_{0+}$) with the electric dipole amplitude
$E_{0+}$ calculated to order $q^2$ from the chiral effective Lagrangian. A
welcome feature of such an approach is that it does not introduce any further 
adjustable parameter. Consequently  meson-baryon interactions
and meson photoproduction are strongly tied together and the fits of e.g. the 
finite range parameters are controlled by both sets of data. For the
description of the photoproduction reactions $\gamma p \to \eta p,\,
K^+\Lambda,\, K^+ \Sigma^0,\, K^0 \Sigma^+$ we have to know the photoproduction
potentials $B_{0+}$ for $\gamma p \to \phi B$, where $\phi B$ refers to the
meson-baryon states with total isospin $I=1/2$ or $I=3/2$ and isospin
projection $I_3=+1/2$. We label these states by an index which runs from 1 to
6, which refers to $|\pi N\rangle^{(1/2)}$, $|\eta N\rangle^{(1/2)}$, $|K
\Lambda\rangle^{(1/2)}$, $|K \Sigma \rangle^{(1/2)}$, $|\pi N\rangle^{(3/2)}$
and  $|K\Sigma\rangle^{(3/2)}$, in that order. The resulting expressions
involve as parameters only the axial vector coupling constants $F$ and $D$ and
read    
\begin{eqnarray}
B_{0+}^{(1)} &=& {e M_N \over 8 \pi f \sqrt{3s}} ( D+F)\,( 2 X_\pi +
Y_\pi)\,\,, \nonumber \\
B_{0+}^{(2)} &=& {e M_N \over 8 \pi f \sqrt{3s}}(3F-D)\,Y_\eta\,\,,\nonumber \\
B_{0+}^{(3)} &=& {e \sqrt{M_N M_\Lambda} \over 8 \pi f \sqrt{3s}}(-D-3F)X_K
\,\,, \nonumber \\
B_{0+}^{(4)} &=& {e \sqrt{M_NM_\Sigma} \over 8 \pi f \sqrt{3s}} ( D-F) (X_K + 2
Y_K) \nonumber \,\,,\\
B_{0+}^{(5)} &=& {e \sqrt{2}M_N \over 8 \pi f \sqrt{3s}}( D+F)(Y_\pi-X_\pi)
\,\,, \nonumber \\
B_{0+}^{(6)} &=&{e\sqrt{2M_NM_\Sigma} \over 8 \pi
f\sqrt{3s}}(D-F)(X_K-Y_K)\,\,, \end{eqnarray}
where $X_\phi$ and $Y_{\phi}$ are dimensionless functions depending on the
center of mass energy $E_\phi$ and the mass $m_\phi$ of the photoproduced
meson. $X_\phi$ takes the form
\begin{equation}
X_\phi  = {1\over 2} - {1\over 4 M_0}\biggl( 2E_\phi +{m_\phi^2 \over
E_\phi}\biggr) + \biggl(1+{m_\phi^2 \over 2 M_0 E_\phi}\biggr){m_\phi^2 \over 2
E_\phi \sqrt{E_\phi^2 -m_\phi^2}} \ln{E_\phi+\sqrt{E_\phi^2-m_\phi^2}\over
m_\phi} \,\,,\end{equation}
and it sums up the contributions of all tree diagrams to the s-wave
photoproduction multipole of a positively charged meson. The logarithmic term
comes from the meson pole diagram in which the photon couples to the positively
charged meson, and its analytic continuation below threshold ($E_\phi <
m_\phi$) is done via the formula
\begin{equation}
{\ln(x+\sqrt{x^2-1}) \over \sqrt{x^2-1}}  = { \arccos x\over \sqrt{1-x^2}}
\,\,. \end{equation}
If the photoproduced meson is neutral  the corresponding sum of diagrams
leads to a simpler expression,
\begin{equation}
Y_\phi = -{1\over 3M_0} \biggl( 2 E_\phi + {m_\phi^2 \over E_\phi}\biggr)\,\,,
\end{equation}  
for the reduced s-wave multipole. Infinitely many rescatterings of the
photoproduced meson-baryon state due to the strong interaction are summed up
via the Lippmann-Schwinger equation. This is shown graphically in Fig.1.  The
"full" electric dipole amplitude $E_{0+}^{(i)}$ for channel $i$ is then given
by 
\begin{equation}
E_{0+}^{(i)} = \sum_j [(1-V\cdot G)^{-1}]_{ij}\,B_{0+}^{(j)}\,\,,\end{equation}
where $V$ is the matrix of the strong interaction potential and $G$ the
diagonal propagator matrix defined in eq.(17). We note that the "full" $E_{0+}$
amplitudes fullfil Watson's final state theorem, i.e. the  phase of the complex
number $E_{0+}$ is equal to the strong interaction phase (in this simple form
the theorem applies only below the $\eta N$ threshold where just one channel is
open). From $E_{0+}^{(i)}$ one can finally compute the total (s-wave)  
photoproduction cross section for the meson-baryon final state $i$,
\begin{equation} \sigma^{(i)}_{\rm tot} = 4 \pi {k_i \over k_\gamma}
|E_{0+}^{(i)}|^2\,\,,  \end{equation}
with $k_\gamma = (s-M_N^2)/2 \sqrt{s}$ the photon center of mass energy and $s=
M_N^2+2M_N E_\gamma^{lab}$ in terms of the photon lab energy $E_\gamma^{lab}$.

We also calculate within the present framework the cross section
for $\eta$-photoproduction from neutrons ($\gamma n\to \eta n$). For this
purpose we have to know the photoproduction potentials for $\gamma n \to \phi
B$, where $\phi B$ is a meson-baryon state with total isospin $I=1/2$ and third
component $I_3=-1/2$. The respective potentials are distinguished by a tilde
from the previous ones and read 
\begin{eqnarray}
\tilde B_{0+}^{(1)} &=& {e M_N \over 4 \pi f \sqrt{3s}} ( D+F)\,( Y_\pi-X_\pi)
\,\,,\nonumber \\
 \tilde B_{0+}^{(2)} &=& \tilde B_{0+}^{(3)} =0 \,\,,\nonumber \\
\tilde B_{0+}^{(4)} &=& {e \sqrt{M_NM_\Sigma} \over 4 \pi f \sqrt{3s}} ( D-F)
(X_K - Y_K)\,\,. \end{eqnarray}
The reason for the second ($\gamma n \to \eta n$) and third ($\gamma n \to K^0
\Lambda)$ neutron photoproduction potential being zero is that here the final
state involves a neutral baryon and a neutral meson to which the
photon cannot couple directly (via the charge). Thus the s-wave meson
photoproduction amplitude vanishes to order $q^2$ in these channels. We will 
see later that the $K\Lambda $ channel is very important for $\eta$-production
off the proton and therefore the order $q^2$ result $\tilde B_{0+}^{(3)}=0$
will lead to a too strong reduction of the $\gamma n \to \eta n$ cross
section. To cure this problem, we include for these double neutral channels
$(\eta n,\,K^0 \Lambda)$ the first correction arising from the coupling of the
photon to the neutral baryon via the anomalous magnetic moment 
\begin{eqnarray} \delta \tilde B_{0+}^{(2)} &=& {e M_N\,\kappa_n \over 48 \pi f
M_0^2\sqrt{3s}} (3F-D) ( 4E_\eta^2-m_\eta^2) \,\,,\nonumber \\
\delta \tilde B_{0+}^{(3)} &=&{e\sqrt{M_NM_\Lambda} \over 96 \pi f M_0^2
\sqrt{3s}}(D+3F) \big[ \kappa_\Lambda (2m_K^2- 5E_K^2)-3 \kappa_n E_K^2   
\bigr] \,\,, \end{eqnarray}
where $\kappa_n = -1.913$ and $\kappa_\Lambda = -0.613$ are the anomalous 
magnetic moments of the neutron and the $\Lambda$-hyperon. 

The extension of our formalism to meson electroproduction (in s-waves) is
straightforward. In electroproduction one has to consider two s-wave
multipoles, the transverse one $E_{0+}$ and the longitudinal one $L_{0+}$, 
which furthermore depend on the virtual photon momentum transfer $q^2<0$. All
steps previously mentioned to construct the s-wave multipole $E_{0+}$ apply to
the longitudinal $L_{0+}$ as well. One only has to generalize the functions
$X_\phi$ and $Y_\phi$ to a transverse and a longitudinal version, which
furthermore  depend on the virtual photon momentum transfer $q^2<0$. The
corresponding somewhat lengthy formulae for $X_\phi^{\rm trans}$ and
$X_\phi^{\rm long}$ can be found in the appendix whereas the
$Y_\phi$-functions do not change, $Y_\phi^{\rm trans}= Y_\phi^{\rm long} =
Y_\phi$, with $Y_\phi$ given in eq.(23).  This completes the discussion of the
formalism necessary to describe meson photo- and electroproduction within our
coupled channel approach.  

\section{Results}
First we have to fix the parameters. For the six channels involved in $K^-p$
scattering we work, as in \cite{ksw1}, in the particle basis taking into
account isospin breaking in the baryon and meson masses but use potentials
$C_{ij}$ calculated in the isospin limit. Then the $K^-p$ and $\overline{K^0}n$
threshold are split and  cusps at the $\overline{K^0}n$ threshold become 
visible in the cross sections. In this six channel problem we allow for three
adjustable range parameters $\alpha_{\overline K N},\, \alpha_{\pi \Lambda}$
and  $\alpha_{\pi \Sigma}$. For the coupled ($\pi N,\,\eta N,\, K \Lambda,\,
K\Sigma$) system we work in the isospin basis as mentioned in section II.B and 
use masses $m_\pi = 139.57$ MeV, $m_K= 493.65$ MeV, $m_\eta=547.45$ MeV,
$M_N=938.27$ MeV, $M_\Lambda= 1115.63$ MeV and $M_{\Sigma} = 1192.55$ MeV, a
choice which averages out most isospin breaking effects. Here we allow for four
adjustable range parameters $\alpha_{\pi N},\,\alpha_{\eta N},\,\alpha_{K
\Lambda}$ and $\alpha_{K\Sigma}$. These seven range parameters and the two
unconstrained combinations of $d$-parameters (in the chiral Lagrangian) were
fixed in a best fit to the data discussed below. We also allowed for optimizing
the parameters $M_0,\, b_0$ and $D$ within narrow ranges. The best fit gave for
the latter $D=0.782$, $M_0=1054$ MeV and $b_0= -0.3036$ GeV$^{-1}$. The last
number, together with the known $b_D,b_F$ leads to  
\begin{equation} \sigma_{\pi N} = 29.4 \,{\rm MeV}\,, \quad y = 0.065\,,
\quad \sigma_{Kp} = 231.6 \, {\rm MeV} \end{equation}
Clearly, as expected the $\pi N$-sigma term is too small if $y$ (the
strangeness content of the proton is small) and the $Kp$-sigma term is in 
reasonable
agreement with other estimates. For the other Lagrangian parameters we find
$d_0 = -0.9189$ GeV$^{-1},\,d_D =0.3351$ GeV$^{-1},\,d_F =-0.4004$
GeV$^{-1},\,d_1 =-0.0094$ GeV$^{-1}$, subject to the two constraints
eqs.(11,12). Note that these numbers are not directly comparable to those in
\cite{ksw1,ksw2}, since here the linear dependent $d_2$-term has been
eliminated. Instead one has to  compare with $d_0+d_2/2,\, d_D-d_2,\,d_F,\,d_1
+d_2$. The best fit of the range parameters gives $\alpha_{\overline KN}= 724$
MeV, $\alpha_{\pi\Lambda}= 1131$ MeV, $\alpha_{\pi \Sigma}= 200$ MeV and
$\alpha_{\pi N}= 522$ MeV, $\alpha_{\eta N}= 665$ MeV, $\alpha_{K \Lambda}=
1493$ MeV, $\alpha_{K\Sigma}= 892$ MeV. (We give sufficiently many digits here
in order to make the numerical results reproducible). On sees that most of the
range parameters are indeed in the physically expected two-pion to vector meson
mass range. We also note that the best fit is very rigid and does not allow
e.g. for a 5$\%$ deviation of the parameters from the values quoted above. Let
us now discuss the fits to the data in detail. 

\subsection{$K^-p$ Scattering}

Fig.2 shows the results for the six $K^-p$ elastic and inelastic channels
$K^-p \to K^-p,$ $\overline{K^0}n$, $\pi^0 \Lambda,$ $\pi^+\Sigma^-$, $\pi^0
\Sigma^0$, $\pi^- \Sigma^+$. As in \cite{ksw1} one finds good agreement with 
the available low energy data below 200 MeV kaon lab momentum. We present
these results here just to make sure again that indeed a large amount of data
can be fitted simultaneously. For the threshold branching ratios  $\gamma, 
R_c, R_n$ (defined in eq.(21) of \cite{ksw1}), we find here $\gamma =
2.33\, (2.36\pm0.04),\, R_c= 0.65\, (0.66\pm0.01),\,R_n=-0.23\,(0.19\pm 0.02)$,
where the numbers given in brackets are the empirical values. In Fig.2 one 
observes cusps in the cross  sections at the $\overline{K^0}n$ threshold, which
are a consequence of unitarity and the opening of a new channel.  Unfortunately
the existing data are not precise enough to confirm this structure. The status 
of the  $K^-p$ scattering  data will improve with DA$\Phi$NE at Frascati
producing intense kaon beams at 127 MeV lab momentum, and in particular once 
the planned kaon facility at KEK will become available. Fig.3 shows real and 
imaginary parts of the calculated total isospin $I=0$ $\overline KN$ s-wave
scattering amplitude in the region $1.35$ GeV $<\sqrt s < 1.45 $ GeV. The
resonance structure around 1405 MeV is clearly visible. In our framework it is
due to the formation of a quasi-bound $\bar K N$-state. It can decay into $\pi
\Sigma$ and thus receives its width of about 25 MeV. We want to stress here
that no effort has been made to prescribe the position or width of the
$\Lambda(1405)$ below the $K^-p$ threshold. Fig.3 results by merely fitting
the scattering data in Fig.1 above the $K^-p$ threshold (together with many
other data).   

\subsection{Eta and Kaon Photoproduction}

The most precise data available are those for $\eta$-photoproduction off 
protons $(\gamma p \to \eta p)$ taken at MAMI \cite{krusche} from threshold at
707 up to 800 MeV photon lab energy. These 54 data points of the total cross
section (full circles) have the highest statistical weight in our fit and they
can indeed be perfectly reproduced as seen in Fig.4. Note that the measured
angular distribution of the $\gamma p \to \eta p$ differential cross section
\cite{krusche} are almost isotropic, thus one can safely identify the total 
cross section with the s-wave cross section.  For the ELSA data \cite{schoch}
(open squares) from 800 to 900 MeV photon lab energy one also finds very good
agreement. Angular distributions have not been published for this case, so we 
assume s-wave dominance. Note that the ELSA data stem from electroproduction
with a very low virtual photon momentum transfer of $q^2=-0.056$ GeV$^2$. We
have actually checked that the corrections from the nonzero $q^2$ and the
longitudinal s-wave $L_{0+}$ to the total cross section
\begin{equation} \sigma_{\rm tot} = 4 \pi {k_\eta \over k_\gamma} (
|E_{0+}|^2 + \epsilon_L |L_{0+}|^2) \end{equation} 
with $\epsilon_L= - 4\,\epsilon  \, s\, q^2(s-M_N^2+q^2)^{-2}$ and $\epsilon =
0.78$ (virtual photon polarization at the ELSA-experiment) are negligible (less
than $6\%$) within error bars. For the
s-wave multipole at the $\eta p$ threshold we get
\begin{equation} E_{0+}^{thr}(\gamma p \to \eta p) = (7.62+ 14.12\,i) \cdot
10^{-3} m_\pi^{-1} \,\,.\end{equation}
This number decomposes as follows. The potentials $B_{0+}^{(j)}$ have the
values $3.65$, $-2.25$, $-15.97$ and $0.78$ (in units of $10^{-3}m_\pi^{-1}$) 
for the $\pi N$, $\eta N$, $K\Lambda$, $K\Sigma$ channels, respectively. The
strong interaction part given by the matrix elements $[(1-V\cdot G)^{-1}]_{2j}$
in eq.(24) multiply these values by complex numbers $(-0.48+0.54i)$, $(1.35+
0.39i)$, $(-0.81-0.82i)$ and $(-0.71-0.08i)$, respectively. Interestingly, it  
is the initial photoexcitation of the $K^+\Lambda$ state which after infinitely
many rescatterings to $\eta p$ makes the largest contribution. The absolute
value $|E_{0+}^{thr}|= 16.04\cdot 10^{-3}m_\pi^{-1}$ is in good  agreement with
determinations in models having an explicit $S_{11}(1535)$ resonance. Our
ratio of imaginary part to real part is 1.85, somewhat larger than the typical
values around 1 found in resonance model fits \cite{krusche}. However, since 
$\eta$-photoproduction  is totally s-wave dominated  in the first 100 MeV above
threshold, the real and imaginary parts cannot be disentangled experimentally.

Fig.4 also shows the  calculated cross sections for $\gamma n \to \eta n$
together with present (preliminary) extractions from the incoherent
$\eta$-photoproduction on the deuteron \cite{kruschepriv}. According to these 
the $\eta$-photoproduction total cross section from the neutron is 2/3 of the
proton one with an uncertainty of $\pm 10\%$. Our calculation including the
anomalous magnetic moment pieces $\delta \tilde B_{0+}^{(2,3)}$ gives a
reasonable description of these data. However, the neutron to proton ratio
is not just a constant 2/3, but shows a stronger energy dependence. Finally, in
Fig.5 we show the ratio of longitudinal and transverse s-wave multipoles
$|L_{0+}/E_{0+}|$ for $\eta$-electroproduction from protons versus the virtual
photon momentum transfer $q^2<0$. The chosen total center of mass energy is
$\sqrt s=1.54$ GeV, where the cross section is expected to be maximal. The
ratio has a value of $0.47$ at the photon point $q^2=0$ and decreases 
rapidly, being close to zero at $q^2 =-0.54$ GeV$^2$. In future
$\eta$-electroproduction experiments at ELSA or CEBAF the longitudinal to
transverse ratio $|L_{0+}/E_{0+}|$ could be measured and it will be
interesting to see whether there is indeed a depletion around $q^2 \simeq -0.5$
GeV$^2$.

In Fig.6 we present results for kaon photoproduction, $\gamma p \to K^+\Lambda,
\, K^+ \Sigma^0, K^0 \Sigma^+$. In these channels the data are more sparse in
the region up to 200 MeV above threshold and the statistical weight of these
data in our combined fit is therefore low. Nevertheless one finds a good 
description of the first few data points in each channel. From the measured
angular distributions \cite{bockhorst,goers} only the first data point for
$\gamma p \to K^0 \Sigma^+$ and the first four for $\gamma p \to K^+\Lambda$
and $\gamma p \to K^+ \Sigma^0$ can be considered as s-wave dominated.
Interestingly, the $\gamma p\to K^+ \Lambda$ s-wave total cross section shows a
strong cusp at the opening of the $K\Sigma$-threshold, which again is a 
consequence of unitarity and multi-channel dynamics. The present $\gamma p\to
K^+ \Lambda$ data do not clearly show such a cusp and it may be covered by the
p-waves which become sizeable above $E_\gamma = 1.1$ GeV. Interestingly if one
divides out of the total cross sections the two-body phase space, $|E_{0+}(
\gamma p \to K^+ \Lambda)|^2$ indeed shows a maximum around $E _\gamma =1046$
MeV, the $K\Sigma$ threshold.  Only more data allowing for a multipole
decomposition can clarify whether a strong cusp is present in the $\gamma p \to
K^+ \Lambda$ s-wave multipole at the $K \Sigma$-threshold.  Note that the
analogous pion induced reaction $\pi^- p \to K^0 \Lambda$ (see Fig.7) shows
this $K\Sigma$-cusp more clearly. The threshold value of the s-wave multipole
is $E_{0+}^{thr}(\gamma p \to K \Lambda) = (-1.78-3.50\,i) \cdot 10^{-3} m_\pi^
{-1}$ and at the cusp one finds $E_{0+}^{K\Sigma-thr}(\gamma p \to K \Lambda) =
(-4.06-2.36\,i) \cdot 10^{-3} m_\pi^{-1}$.  The first two data points for
$\gamma p \to K^+ \Sigma^0$ are somewhat overshot (see Fig.6) and we find 
$E_{0+}^{thr}(\gamma p\to K^+ \Sigma^0)=(4.15+3.11\,i) \cdot 10^{-3}m_\pi^{-1}$
for the s-wave multipole at threshold, which may be too large in magnitude. One
finds however that the first few data points for $\gamma p \to K^+ \Sigma^0$ do
not follow the phase space. Again, more data and a multipole decomposition are
needed here. Finally, we show in Fig.6 the $\gamma p \to K^0 \Sigma^+$
channel. This curve is really a prediction with no data included in the fit
and, interestingly, the first data point of the recent analysis \cite{goers} is
reproduced. The data points further above threshold are not comparable to our 
s-wave approximation since the measured angular distributions in $\gamma p \to
K^0 \Sigma^+$ are strongly anisotropic. For the threshold value of $E_{0+}$ we
find $E_{0+}^{thr}(\gamma p \to K^0 \Sigma^+) = (1.34+ 3.38\,i)\cdot 10^{-3}
m_\pi^{-1}$. Note that our approach does not have the problem of overpredicting
the $\gamma p \to K^0 \Sigma^+$ channel, at least in s-wave approximation. 

\subsection{Pion Induced Eta and Kaon Production}

In Fig.7, we show our results for pion induced eta and kaon production. The
cross section data for $\eta$-production $\pi^- p \to \eta n$ are the
selection of \cite{nefkens}, and the kaon production cross sections of
$\pi^-p \to K^0 \Lambda,\, K^0 \Sigma^0, K^+ \Sigma^-$ and $\pi^+ p \to K^+
\Sigma^+$ are taken from the compilation \cite{data}. The agreement of the
coupled channel calculation with the data is very good even for energies
considerably above threshold, except for the (pure isospin 3/2) $\pi^+ p \to
K^+ \Sigma^+$  channel, where the s-wave approximation seems to hold only for
the first data point. Since the pion induced data have quite large error bars 
their statistical weight is low. Therefore the good agreement with the existing
data is a highly non-trivial consistency check of our coupled channel approach
which simultaneously describes strong and electromagnetic meson-baryon
interactions. As a byproduct of this calculation we extract the $\eta N$ s-wave
scattering length for which we find $a_{\eta N} = (0.20+ 0.26\,i)$ fm. Its 
imaginary part is almost fixed by unitarity from the rise of the $\pi^- p \to 
\eta n$ cross section and it is in agreement with \cite{batinic}. However, the
real part of $a_{\eta N}$ is almost a factor of 5 smaller than in the unitary
resonance model calculation of \cite{batinic}. As recently pointed out in 
\cite{birbrair}, $a_0(980)$-meson exchange in the
t-channel can equally well describe the $\pi^- \to \eta n$ total cross
sections, but leads to a negative real part of $a_{\eta N}$. Therefore we do
not consider our small Re\,$a_{\eta N} = 0.20$ fm as unrealistic.     

\subsection{Nature of the $S_{11}(1535)$-resonance}
The peaks of the observed total cross sections in $\gamma p\to \eta p$ (see
Fig.4) and $\pi^- p \to \eta n$ (see Fig.7) at $\sqrt s\simeq 1.54$ and $1.53$
GeV, respectively, suggest the presence of an isospin 1/2 nucleon s-wave
resonance, the $S_{11}(1535)$ . Actually the resonance parameters (mass $M^*$
and width $\Gamma$) are determined from a fit of the data using a Breit-Wigner
parametrization with typical values $M^*= 1.48\dots 1.55$ GeV and $\Gamma
\simeq 200$ MeV \cite{krusche,nefkens}. However the closeness of the $\eta N$
threshold ($M_N+m_\eta = 1486$ MeV) causes peculiar features. The  $\eta N$
partial decay width is so strongly energy dependent that the Breit-Wigner curve
$|2(M^*-\sqrt s)-i\Gamma(\sqrt s)|^{-2}$ decreases  monotonically from the
$\eta N$ threshold onwards without showing a resonance peak. Furthermore,
speed plots (absolute values of derivatives of partial wave amplitudes
with respect to $\sqrt s$) in \cite{speedplots} derived from $\pi N$
dispersion analysis show no structure in the $S_{11}$ partial wave at $\sqrt s
\simeq 1535$ MeV, but only the strong $\eta N$ cusp and the second
$S_{11}(1650)$ resonance. In \cite{denschlag} a calculation of
$\eta$-production has been performed using the analytically solvable Lee-model
and it was found that the data can equally well be reproduced with just a 
strong background instead of a $S_{11}(1535)$ resonance pole. The resonance
poles actually found in this calculation lie on the wrong (fourth)
Riemann-sheet.     

Let us now take a closer look at these issues in the present coupled channel
calculation based on the chiral Lagrangian. As first pointed out in \cite{ksw2}
the chiral $K\Sigma$ isospin $I=1/2$ s-wave interaction is strongly attractive.
It can thus built up a resonance-like state with the properties of the $S_{11}
(1535)$. In Fig.8 we show that eigenphase of the four-channel $S$-matrix which 
below the $\eta N$-threshold (marked by an arrow) joins continuously with the
elastic $S_{11}$ $\pi N$ phase. Its value of $52^\circ$ at the $\eta N$
threshold is somewhat larger than the empirical $40^\circ$ (see Fig.2a in
\cite{ksw2}). The eigenphase passes through $90^\circ$ at $\sqrt s= 1584$ MeV
with a slope that can be translated into a full width of $\Gamma= 198$
MeV. These numbers (in particular the one for the width $\Gamma$) are in good
agreement with those attributed to the $S_{11}(1535)$. However, as one goes up
in energy the phase starts to decrease once it has reached $125^\circ$ at the
$K\Sigma$-threshold.  As required for a clean resonance the phase does not
change by $180^\circ$. The situation is somehow similar to the isospin $I=0$
$\pi \pi$ s-wave at $\sqrt s = 0.88$ GeV  where the phase of $90^\circ$ does 
not correspond to a scalar resonance but to a strong background. 

On the other hand the dynamically generated $\Lambda(1405)$ in the $(\overline
KN,\pi\Sigma)$-system is a clean resonance in this sense. The $I=0$ $\pi
\Sigma$ s-wave phase shift, which passes through $90^\circ$ at $\sqrt s \simeq
1.41$ GeV has moved above $180^\circ$ at the $\overline K N$ threshold. 

Finally, we show in Fig.9 the real and imaginary part of the s-wave multipole
$E_{0+}(\gamma p \to \eta p)$ in the region $1.45\, {\rm GeV} < \sqrt s <1.67$
GeV. The curves show a resonance-like behavior with the real part of $E_{0+}$
passing through zero at $\sqrt s = 1545$ MeV. Obviously, in this restricted
energy range a clean resonance and a strong background are indistinguishable
from each other.  We conclude that within our coupled channel approach, the 
questionable status of the $S_{11}(1535)$ (resonance or a strong background) is
reconfirmed \cite{speedplots,denschlag}.

\section{Summary}
In summary, we have used the SU(3) chiral meson-baryon Lagrangian at
next-to-leading order together with a unitary coupled channel approach to
describe simultaneously a large number of meson-baryon scattering and meson 
photoproduction processes. The extension to photo- and electroproduction is
parameter free (to order $q^2$). By adjusting only 9 parameters (2 constants in
the Lagrangian and 7 finite range parameters) we are able to successfully 
describe a large amount of low energy data. These include elastic and inelastic
$K^-p$ scattering $K^-p \to K^-p,\, \overline{K^0}n,\, \pi^0 \Lambda,\,\pi^+
\Sigma^-, \,\pi^0 \Sigma^0 ,\, \pi^- \Sigma^+$, eta and kaon photoproduction of
protons (and neutrons) $\gamma p \to \eta p,\, K^+\Lambda,\, K^+\Sigma^0,\, K^0
\Sigma^+$ and $\gamma n \to \eta n$ as well as those of the corresponding pion
induced reactions $\pi^- p \to \eta n,\, K^0 \Lambda,\, K^0 \Sigma^0 , \, K^+
\Sigma^-$ and $\pi^+ p \to K^+ \Sigma^+$. We do not introduce any explicit
resonances, but generate the $\Lambda(1405)$ and $S_{11}(1535)$ as
quasi-bound states of $\overline KN$ and $K\Sigma$. We furthermore took a 
closer  look at the nature of the "$S_{11}(1535)$" in this framework and find
that it could well be a strong background instead of a clean resonance. 
Naturally, the next step for improvement (in particular the description of kaon
photoproduction) is the systematic inclusion of the (strong and
electromagnetic) p-waves. Given that the s-wave approximation does very well in
many channels, this is nontrivial since the p-waves have to be rather selective
in the various  channels. It remains to be seen whether the p-waves derived
from the effective chiral Lagrangian can do this job.
 
\bigskip

\centerline{\bf Acknowledgements}

\bigskip

We are grateful to S. Goers, R. Gothe and B. Krusche for valuable information
on the experimental data. We thank S. Goers for providing us with the new
$\gamma p \to K^0 \Sigma^+$ data before publication. 

\bigskip

\centerline{\bf Appendix}

\bigskip

\renewcommand{\theequation}{A\arabic{equation}}
\setcounter{equation}{0}
Here we collect the explicit formulae for the potential coupling strengths
$C_{ij}= C_{ji}$ for the four channel system consisting of the total isospin
$I=1/2$ meson-baryon states $|\pi N\rangle^{(1/2)},\,|\eta N\rangle^{(1/2)},\,
|K\Lambda  \rangle^{(1/2)},\, |K \Sigma \rangle^{(1/2)}$ (labeled by indices 1,
2, 3, 4, respectively) and for the two channel system consisting of the total
isospin $I=3/2$ states $|\pi N\rangle^{(3/2)},\, |K \Sigma\rangle^{(3/2)}$
(labeled by indices 5 and 6). In this basis one has for the physical states,
\begin{eqnarray}
|\pi^-p\rangle &=& \bigl[\sqrt2|\pi N\rangle^{(1/2)}+|\pi N\rangle^{(3/2)}
\bigr]/ \sqrt3\,\,,  \nonumber \\
|K^+ \Sigma^0\rangle &=& \bigl[|K\Sigma \rangle^{(1/2)} + \sqrt 2 |K\Sigma
\rangle^{(3/2)} \bigr]/\sqrt3\,\,,  \nonumber \\
|K^+ \Sigma^-\rangle &=& \bigl[\sqrt 2|K\Sigma \rangle^{(1/2)} +  |K\Sigma
\rangle^{(3/2)} \bigr]/\sqrt3\,\,,  \nonumber \\
|K^0 \Sigma^+ \rangle &=& \bigl[\sqrt 2|K\Sigma\rangle^{(1/2)} - |K\Sigma
\rangle^{(3/2)} \bigr]/\sqrt3 \,\,,\nonumber \\ 
|K^0 \Sigma^0\rangle &=& \bigl[-|K\Sigma \rangle^{(1/2)} + \sqrt 2 |K\Sigma
\rangle^{(3/2)} \bigr]/\sqrt3 \,\,.  \end{eqnarray}

\begin{eqnarray}
C_{11} & = & -E_\pi  + {1 \over {2 M_0}}(m_\pi^2 - E_\pi^2)
     + 2 m_\pi^2(b_D + b_F + 2 b_0) - E_\pi^2 (d_D + d_F + 2 d_0) \nonumber \\ 
& & + {g_A^2\over 4} (3S_{\pi\pi}-U_{\pi\pi})\,\,, \nonumber \\
C_{12} & = &  2 m_\pi^2 (b_D + b_F) - E_\pi E_\eta (d_D + d_F) +{g_A\over
4} (3F-D)(S_{\pi\eta}+U_{\pi\eta}) \,\,,\nonumber \\ 
C_{13} & = &  {3 \over 8} (E_\pi + E_K) + {3 \over {16M_0}}(E_\pi^2-m_\pi^2+
E_K^2-m_K^2) - {1 \over 2}(m_K^2+m_\pi^2)(b_D+3b_F)  \nonumber \\
 &   & + {1\over 2}E_\pi E_K (d_D+3d_F) -{g_A\over 4}(3F+D)S_{\pi K}
 +{D\over 2} (D-F) U_{\pi K} \,\,,\nonumber \\ 
C_{14} & = & -{1 \over 8}(E_\pi+E_K) - {1 \over16 M_0}(E_\pi^2-m_\pi^2+E_K^2-
m_K^2) + {1 \over 2}(b_F -b_D)  (m_\pi^2+m_K^2)  \nonumber \\
 &   &  + {1\over 2}E_\pi E_K (d_D-d_F-2d_1)+{3\over 4} (D^2-F^2)S_{\pi K}
 +({DF\over 2} -{D^2 \over 6} -F^2) U_{\pi K}\,\,, \nonumber \\
C_{22} & = &  {16 \over 3} m_K^2(b_D-b_F+b_0) + 2 m_\pi^2({5 \over 3} b_F
- b_D-{2 \over 3}b_0)+E_\eta^2 (d_F-{5 \over 3}d_D-2d_0)  \nonumber\\ & & 
+{1\over 12} (3F-D)^2 (S_{\eta \eta}+U_{\eta\eta})\,\,,\nonumber \\
C_{23} & = &  {3 \over 8}(E_\eta +E_K ) + {3 \over {16M_0}} (E_K^2-m_K^2+
E_\eta^2-m_\eta^2) + (b_D+3 b_F)({5 \over 6} m_K^2-  {1 \over 2}m_\pi^2) 
\nonumber \\& &  -E_\eta E_K({d_F \over 2} +  {d_D \over 6} + d_1) +({D^2\over 12} -{3\over 4} F^2) S_{K\eta} +{D\over 2}(F+{D\over 3}
) U_{K\eta} \,\,,\nonumber \\
C_{24} & = &   {3 \over 8} (E_\eta + E_K) + {3 \over {16M_0}} (E_K^2-m_K^2+
E_\eta^2-m_\eta^2) + ({5 \over 2}m_K^2-{3 \over 2}m_\pi^2)(b_F-b_D) \nonumber  
\\ & &   + {1\over 2}E_\eta E_K  (d_D-d_F) +{1\over 4}
(3F-D)(D-F)S_{K\eta}+ {D\over 2}(D-F)U_{K\eta} \,\,, \nonumber \\
C_{33} & = &({10 \over 3}b_D+4b_0) m_K^2- E_K^2 (2d_0+{{5d_D} 
\over 3}) +{3\over 4} (F+{D\over 3})^2 S_{KK}+{3\over 4}(F-{D\over 3})^2 U_{KK}
\,\,,\nonumber\\
C_{34} & = &   2m_K^2b_D-E_K^2 d_D +{1\over 4} (3F+D)(F-D)S_{KK}+{g_A\over
4} (3F-D) U_{KK} \nonumber \,\,,\\
C_{44} & = &-E_K -{1 \over {2 M_0}}(E_K^2-m_K^2) + 2m_K^2(b_D-2b_F+2b_0)+ 
E_K^2(2d_F-d_D-2d_0) \nonumber \\ & & +{3\over 4} (D-F)^2 S_{KK}-{g_A^2\over 4}
U_{KK} \,\,,\nonumber \\
C_{55} & = &{E_\pi \over 2} +{1 \over {4 M_0}}(E_\pi^2-m_\pi^2) +
2m_\pi^2(b_D+b_F+2b_0)- E_\pi^2(d_D+d_F+2d_0) +{g_A^2\over 2}  U_{\pi\pi} 
\,\,,\nonumber \\
C_{56} & = & {1\over 4} (E_\pi+E_K) +{1 \over {8 M_0}}(E_\pi^2 -m_\pi^2+ E_K^2
-m_K^2) + (m_\pi^2+m_K^2)(b_D-b_F) \nonumber \\ & & + E_\pi E_K(d_F-d_D-d_1) 
+ ( {F^2 \over 2} -DF -{D^2 \over 6} ) U_{\pi K} \,\,, \\
C_{66} & = & {E_K\over 2} +{1 \over {4 M_0}}(E_K^2-m_K^2) +
2m_K^2(b_D+b_F+2b_0)-  E_K^2(d_F+d_D+2d_0) +{g_A^2\over 2}  U_{KK} \nonumber
\,\,. \end{eqnarray}
The abbreviations $S_{\phi\phi'}$ and $U_{\phi\phi'}$ stand for  
\begin{equation} S_{\phi \phi'} = {E_\phi E_{\phi'} \over 2 M_0} \,, \qquad
U_{\phi \phi'} = {1\over 3M_0} \biggl( 2 m_\phi^2 + 2m_{\phi'}^2 + {m_\phi^2
m_{\phi'}^2 \over E_\phi E_{\phi'}} - {7\over 2} E_\phi E_{\phi'} \biggr)\,\,.
\end{equation} 
The occurrence of $1/E_\phi$-terms in $U_{\phi\phi'}$ reflects the presence of
the short cut singularities in the s-wave meson-baryon amplitudes below
threshold due to u-channel baryon pole diagrams. Unfortunately, the singularity
of the $1/E_K$-pole term at $s= M_\Sigma^2-m_K^2 >(M_N+m_\pi)^2$ lies slightly
above the $\pi N$ threshold and therefore some (strong and electromagnetic) 
s-wave potentials become singular at $\sqrt s = 1086$ MeV. In contrast to a 
proper four dimensional loop integration the used simple Lippmann-Schwinger
equation in s-wave approximation is not able to get rid of this singularity in
the $T$-matrix or $E_{0+}$. This feature prohibits a continuation of the
present coupled channel calculation down to the $\pi N$ threshold.     

The treatment of meson electroproduction requires the generalization of the
$X_\phi$ and $Y_\phi$ functions to a transverse and a longitudinal
version. There is no change to order $q^2$ in the $Y_\phi$ functions,
i.e. $Y^{\rm trans}_\phi  = Y^{\rm long}_\phi  = Y_\phi$. In contrast to this
the $X_\phi$ functions  become more complicated and depend furthermore on the
virtual photon momentum transfer $ q^2 = -2t$ as follows:
\begin{eqnarray}
X_\phi^{\rm trans} &=& {E_\phi^2 +3t \over2( E_\phi^2+2t)} +
{m_\phi^2(E_\phi^2+2t) +t^2 \over 2 (E_\phi^2 + 2t)^{3/2}}\,W_\phi
+ {E_\phi\over M_0} \biggl\{ {(m_\phi^2+2t)(t-E_\phi^2) \over 4( E_\phi^2+2t)
^2}\nonumber \\ & &-{1\over 2}+ {(m_\phi^2+2t)(2m_\phi^2 t-t^2+
m_\phi^2 E_\phi^2-2t E_\phi^2)\over 4(E_\phi^2 + 2t)^{5/2}}\,W_\phi \biggr\}
\,\,,\end{eqnarray}   
\begin{eqnarray}
X_\phi^{\rm long} &=& {E_\phi^2  \over2( E_\phi^2+2t)} +
{t E_\phi^2 \over 2 (E_\phi^2+2t)^{3/2}}\,W_\phi+{E^3_\phi\over M_0} 
\biggl\{ -{1\over 2E_\phi^2} \nonumber \\ & &+{(m_\phi^2+2t)(4m_\phi^2 t
+3t^2+ 2 m_\phi^2 E_\phi^2) \over 4( E_\phi^2+2t)^2(t^2+2t m_\phi^2 + m_\phi^2
E_\phi^2) }+ {(m_\phi^2+2t)(t- E_\phi^2)  \over 4
(E_\phi^2 + 2t)^{5/2}}\,W_\phi \biggr\} \,\,, \end{eqnarray}   
with the abbreviation
\begin{equation} W_\phi= {1\over\sqrt{E_\phi^2-m_\phi^2} } \ln
{E_\phi^2 + t+  \sqrt{(E_\phi^2-m_\phi^2)(E_\phi^2+ 2t) } \over
\sqrt{m_\phi^2(E_\phi^2+2t)+t^2}}\,\,.  \end{equation}
\bigskip

\centerline{\bf Figure Captions}

\medskip

\begin{itemize}
\item[Fig.1] Graphical representation of the Lippmann-Schwinger equation for
s-wave meson photoproduction. The full, broken and wavy lines represent
baryons, mesons and the photon, respectively.

\item[Fig.2] The total cross sections for the six $K^-p$ elastic and inelastic
scattering channels $K^-p \to K^-p, \overline{K^0}n, \pi^0 \Lambda, \pi^- 
\Sigma^+, \pi^0 \Sigma^0, \pi^+ \Sigma^-$ versus the kaon lab momentum. The
data are taken from \cite{data}. 

\item[Fig.3] The $\overline K N$ s-wave amplitude with isospin $I=0$ versus 
the  total center of mass energy $\sqrt{s}$. The dashed and full curve
correspond to the real and imaginary part of the scattering amplitude.
 
\item[Fig.4] The total cross sections for $\eta$-photoproduction off protons
and neutrons versus the photon lab energy $E_\gamma$. Full circles: MAMI data
\cite{krusche}, open squares: ELSA-data \cite{schoch}. The preliminary neutron
data are 2/3 of the proton cross sections with an error bar of $\pm 10\%$
\cite{kruschepriv}.

\item[Fig.5] The ratio of longitudinal and transverse s-wave amplitudes
$|L_{0+}/E_{0+}| $ for $\eta$-electroproduction off protons at $\sqrt s = 1.54$
GeV versus the virtual photon momentum transfer $q^2<0$.

\item[Fig.6] The total cross sections for kaon photoproduction from protons
$\gamma p \to K^+ \Lambda$, $K^+ \Sigma^0$, $K^0 \Sigma^+$ versus the photon
lab energy $E_\gamma$. The $K^+$-data are taken from \cite{bockhorst}. The
preliminary $K^0\Sigma^+$ data point has been provided by \cite{goers}. 

\item[Fig.7] The total cross sections for pion induced eta and kaon production
$\pi^- p \to \eta n,K^0 \Lambda,\\K^0 \Sigma^0,K^+ \Sigma^-$ and $\pi^+ p
\to K^+ \Sigma^+$  versus the pion lab momentum $p_{lab}$. The
$\eta$-production data are taken from the selection of \cite{nefkens} and the
kaon production data from the compilation \cite{data}.  

\item[Fig.8] The "resonant" eigenphase of the four-channel S-matrix of the
s-wave, isospin $I=1/2$ $(\pi N,\eta N,K\Lambda,K \Sigma)$-system versus the
total center of mass energy $\sqrt s$.

\item[Fig.9] Real (dashed line) and imaginary part (full line) of the s-wave
electric dipole amplitude $E_{0+}(\gamma p \to \eta p)$ versus the total center
of mass energy $\sqrt s$.    
\end{itemize}

\bild{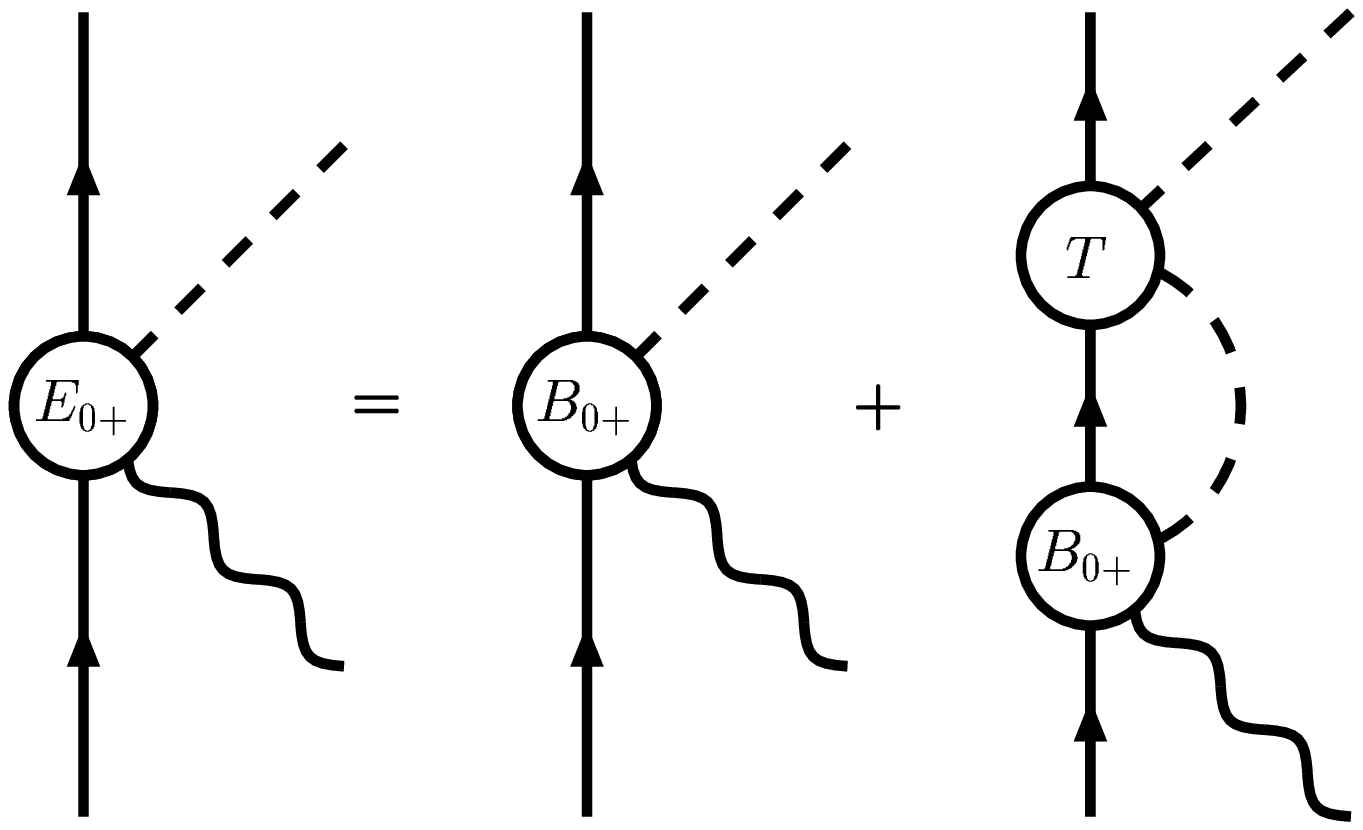}{13}
\vspace*{-9cm}
\begin{center}
\large Figure 1
\end{center}

\newpage

\vspace*{2cm}
\begin{minipage}{7cm}
\bild{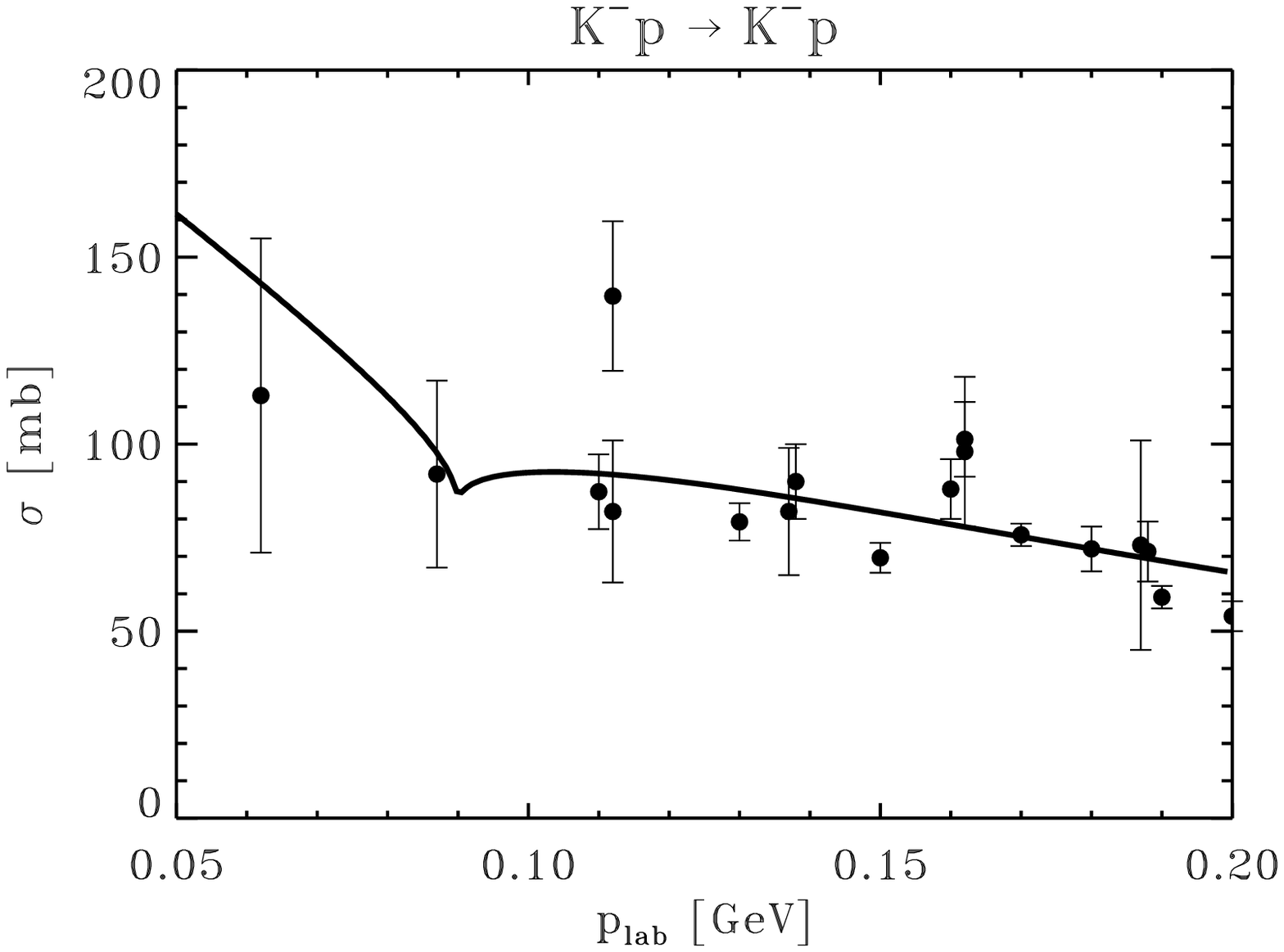}{7.5}
\bild{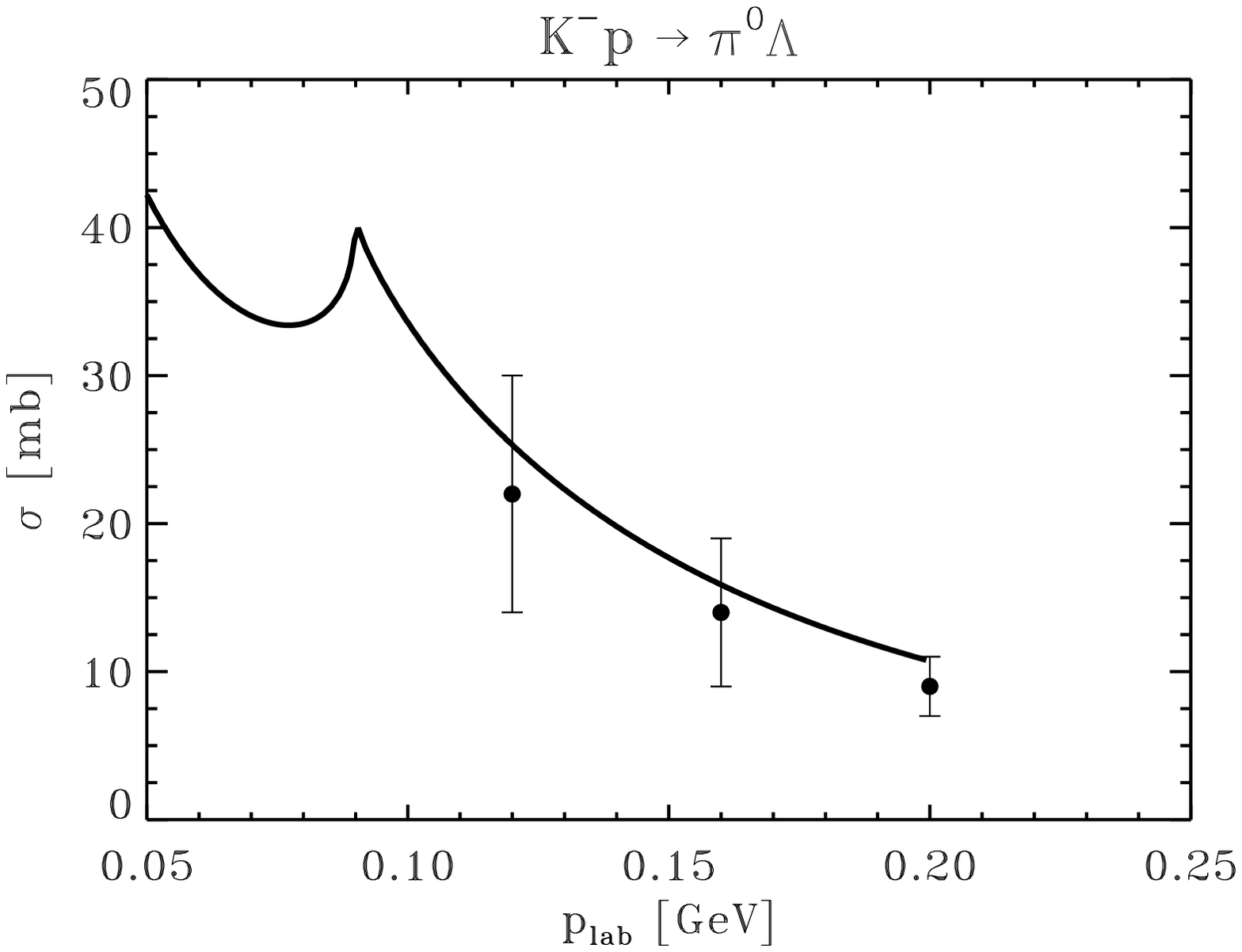}{7.5}
\bild{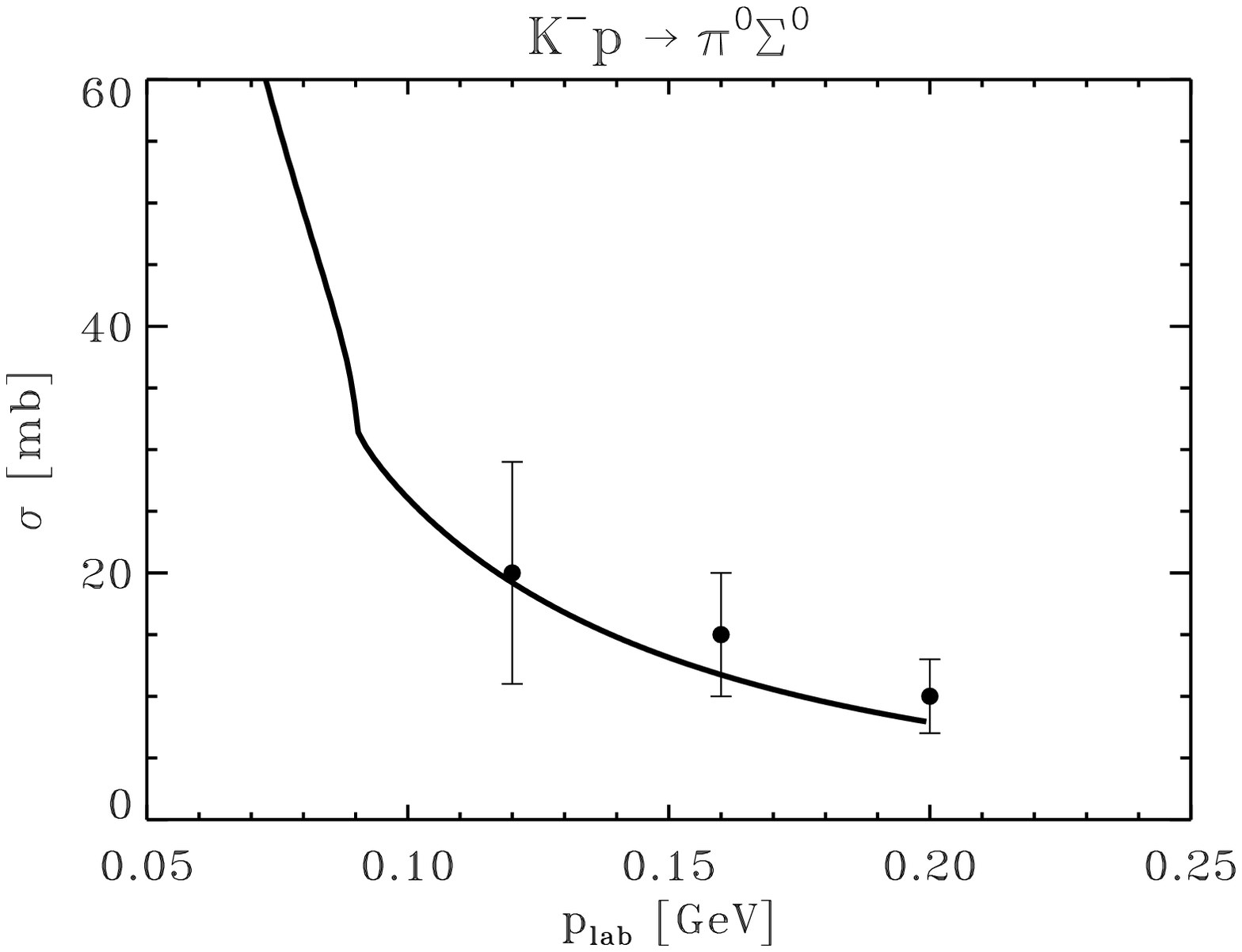}{7.5}
\end{minipage}
\begin{minipage}{7cm}
\bild{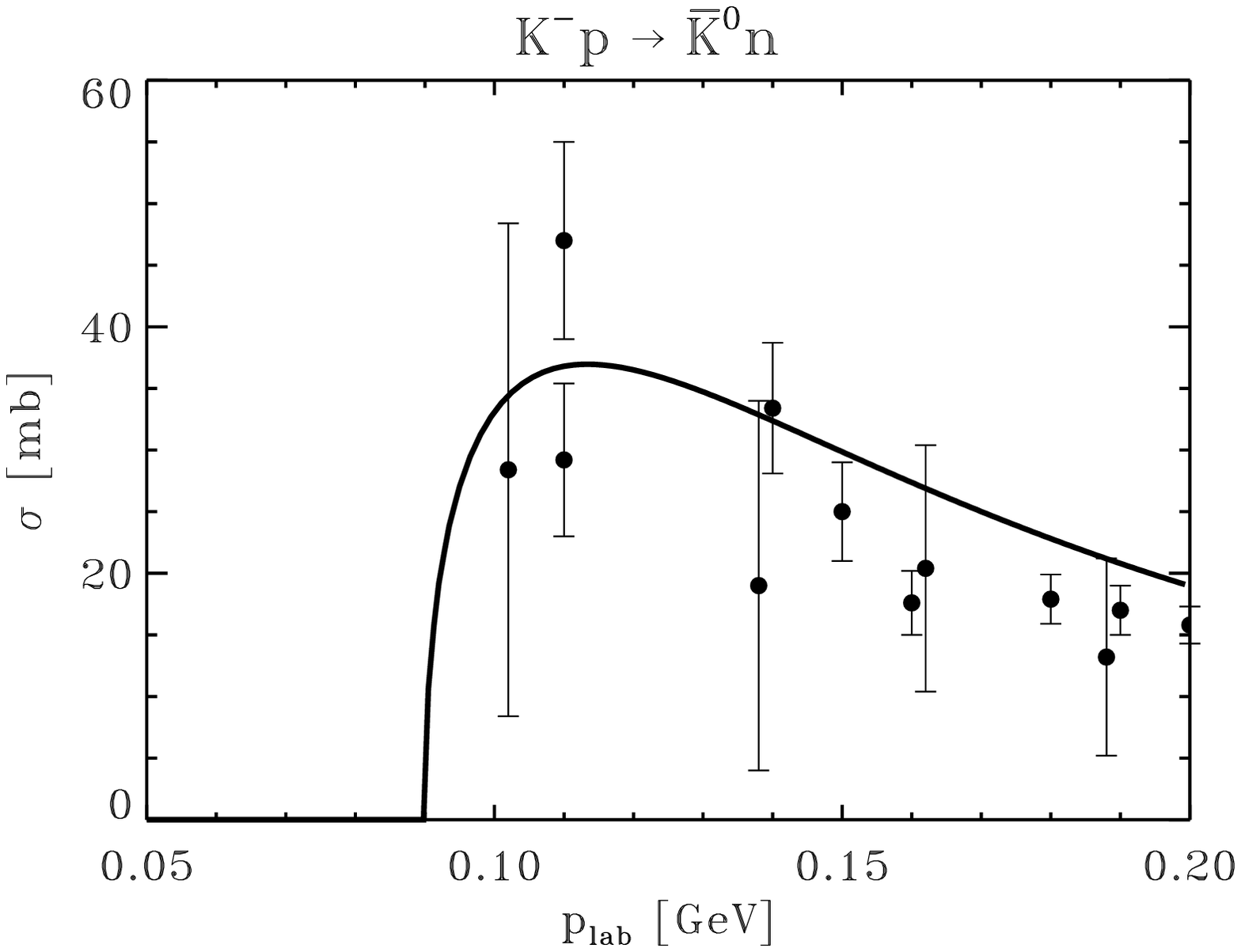}{7.5}
\bild{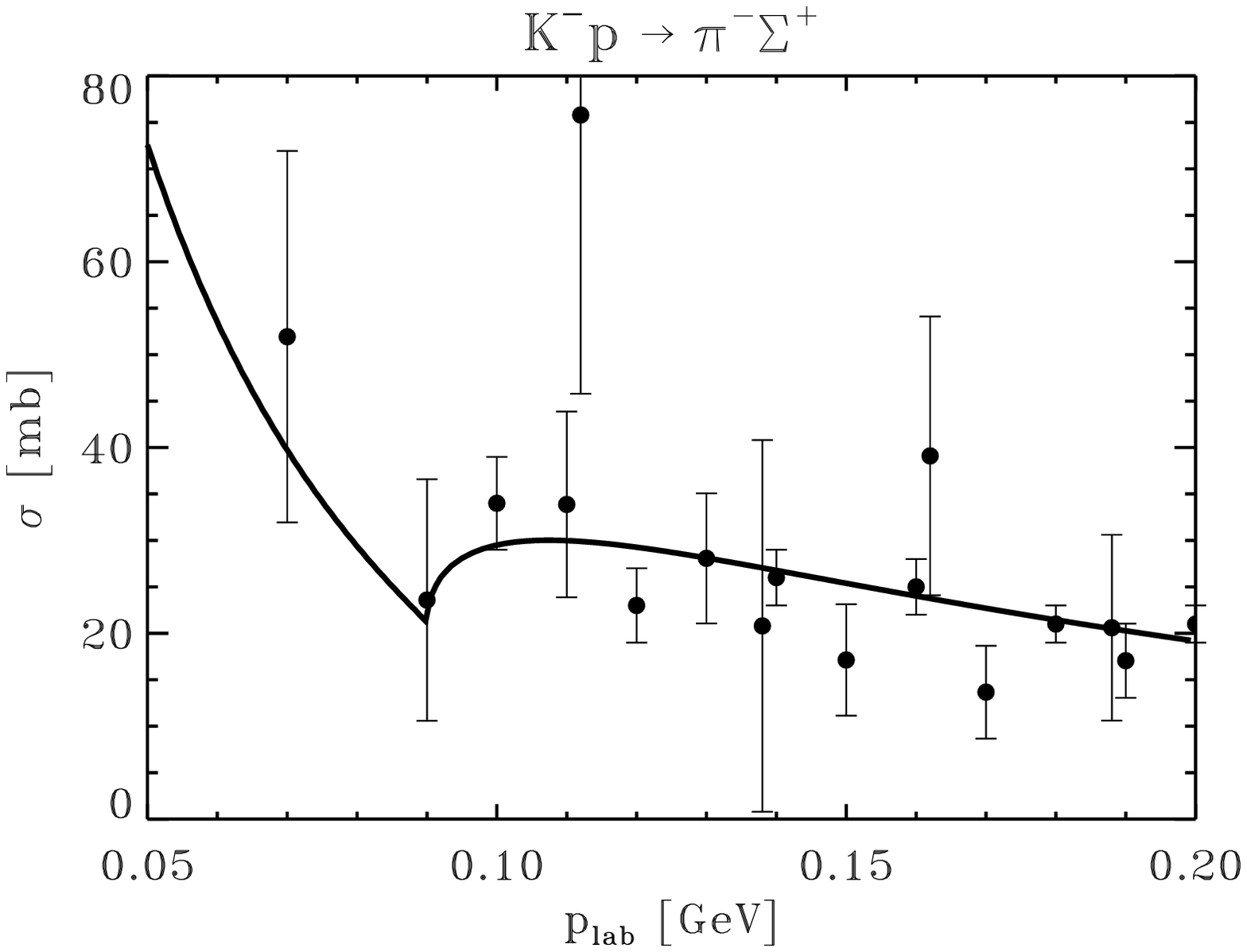}{7.5}
\bild{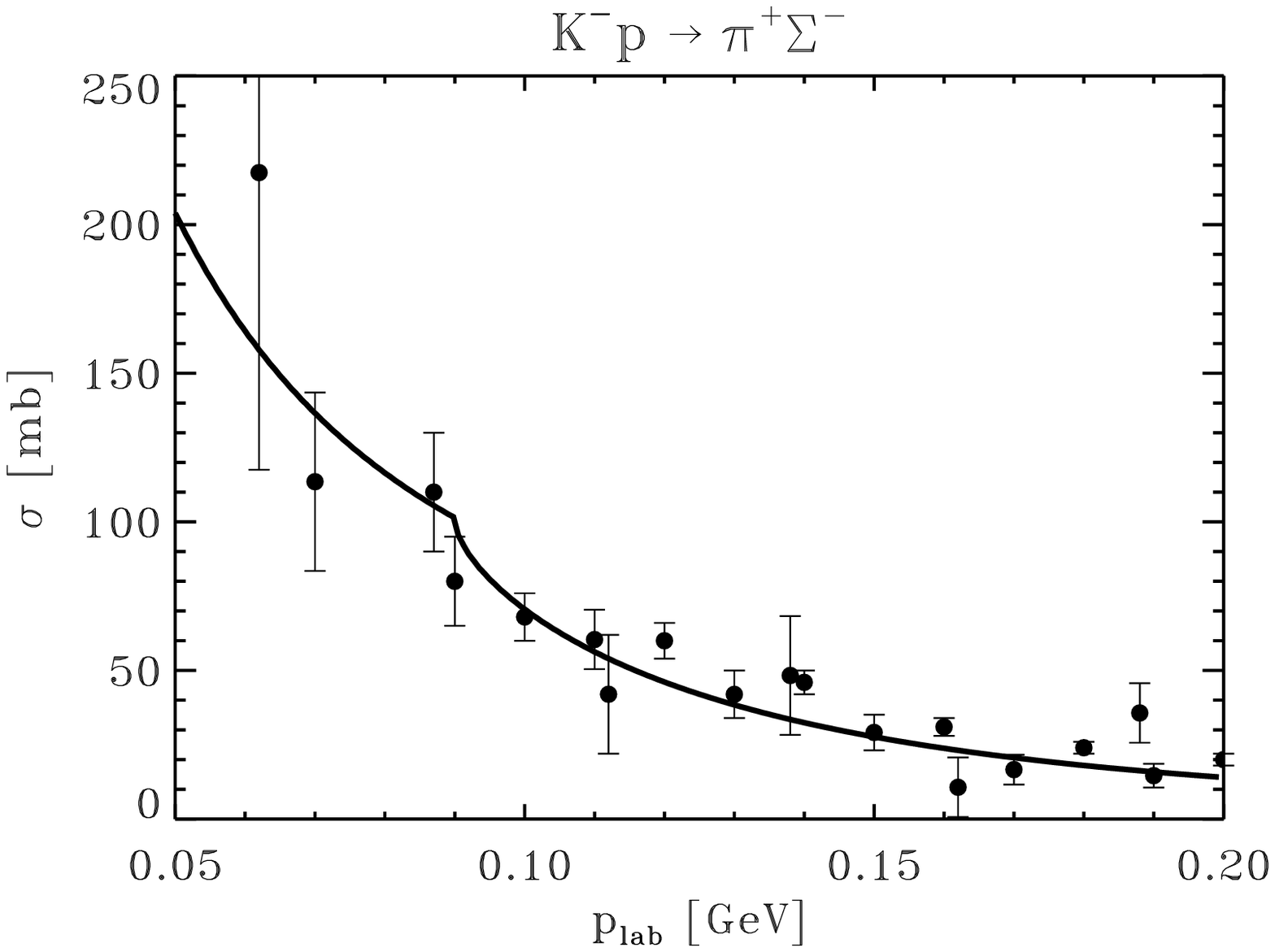}{7.5}
\end{minipage}

\vspace*{1cm}
\centerline{\large Figure 2}

\newpage

\bild{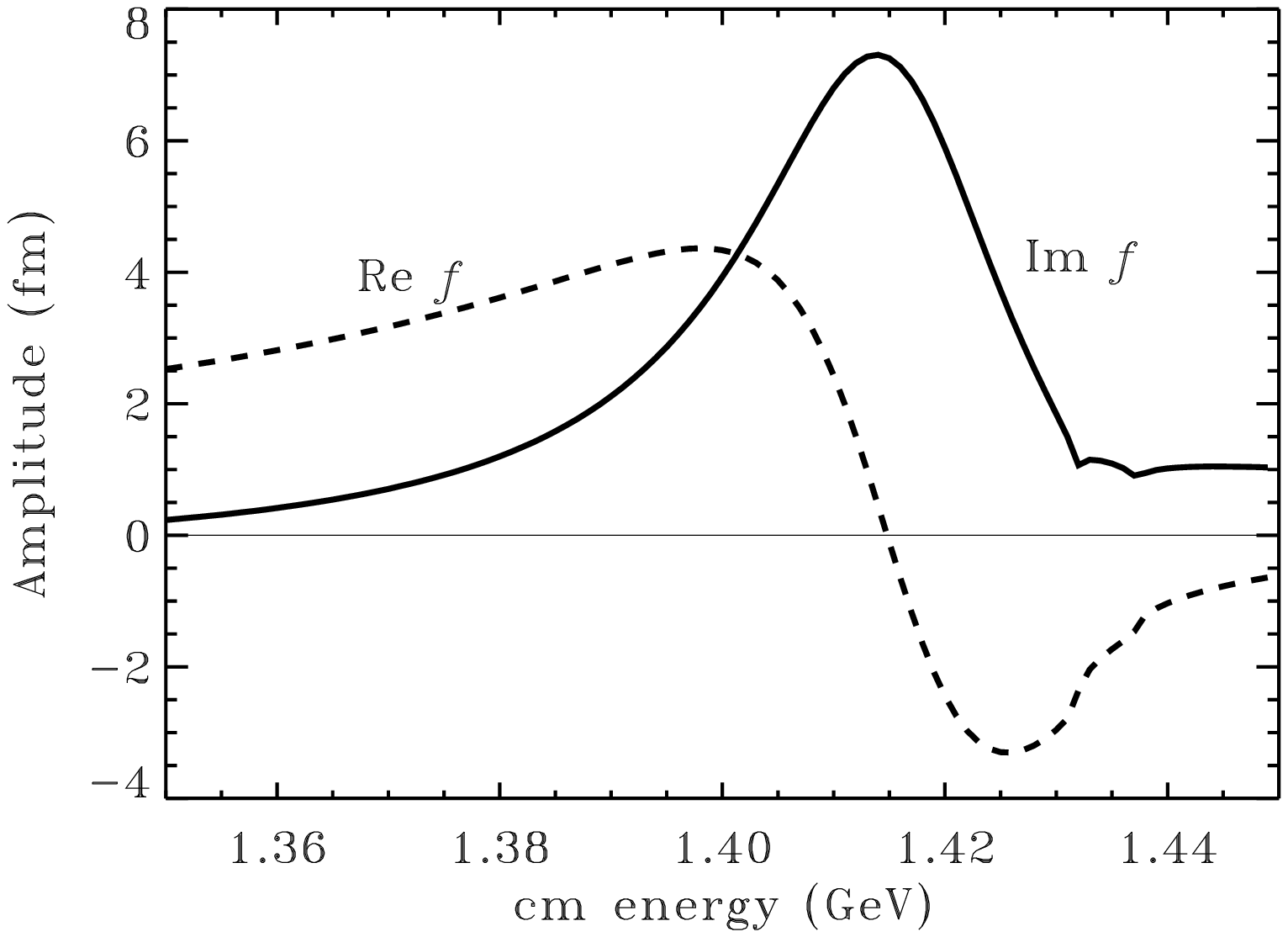}{16}
\vspace*{.5cm}
\centerline{\large Figure 3}

\newpage

\bild{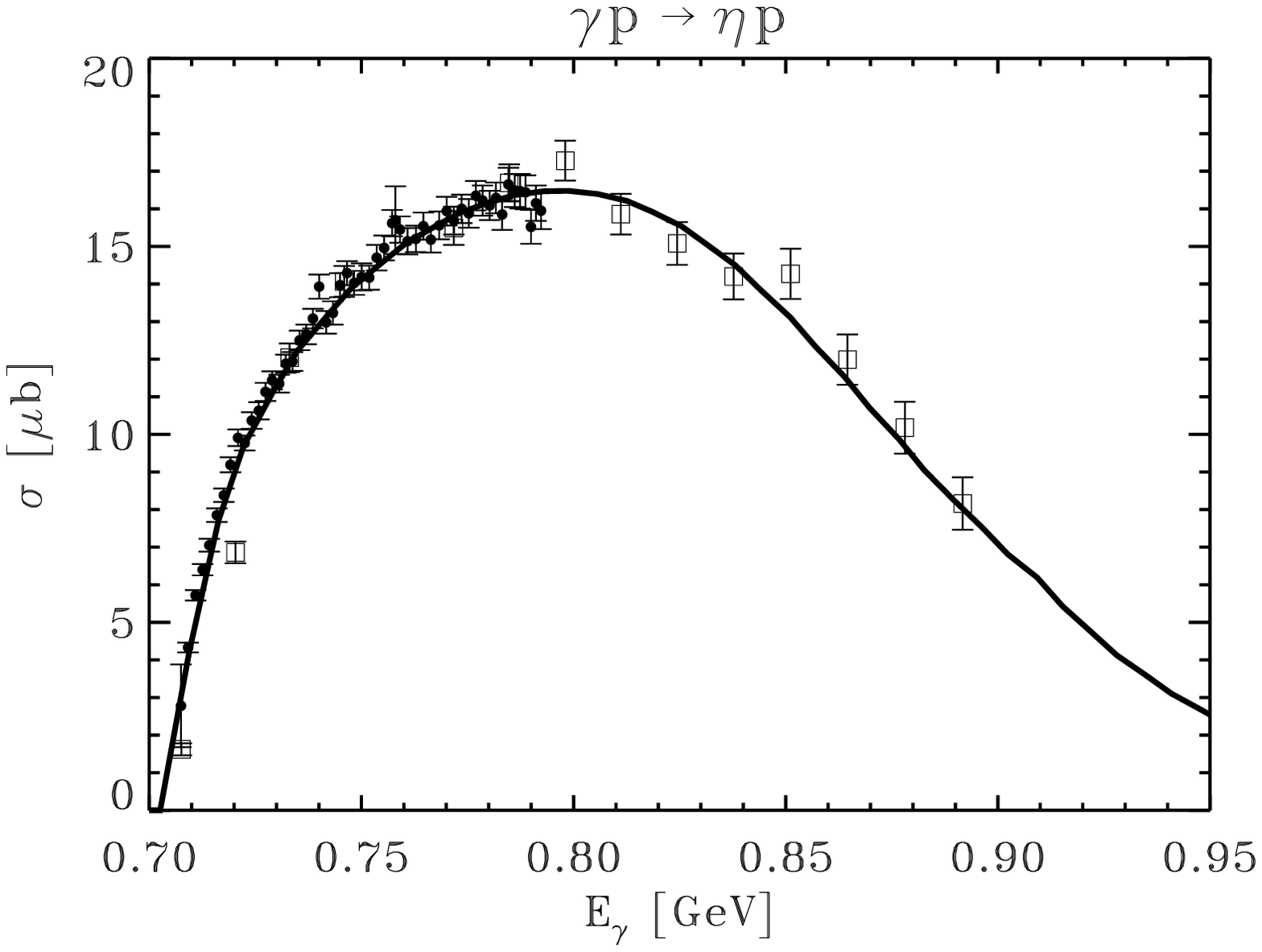}{15}
\bild{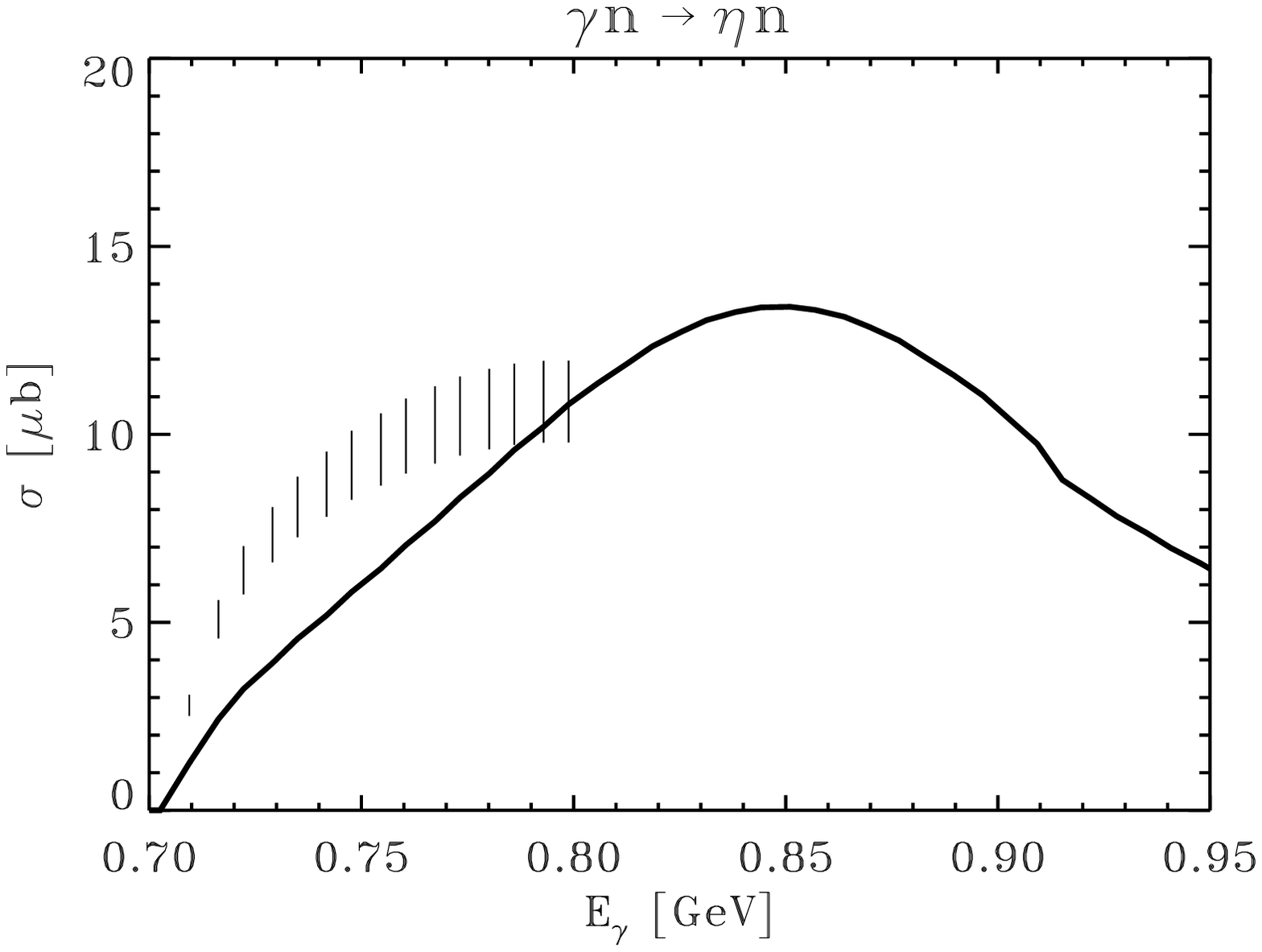}{15}
\vspace*{.5cm}
\centerline{\large Figure 4}

\newpage

\bild{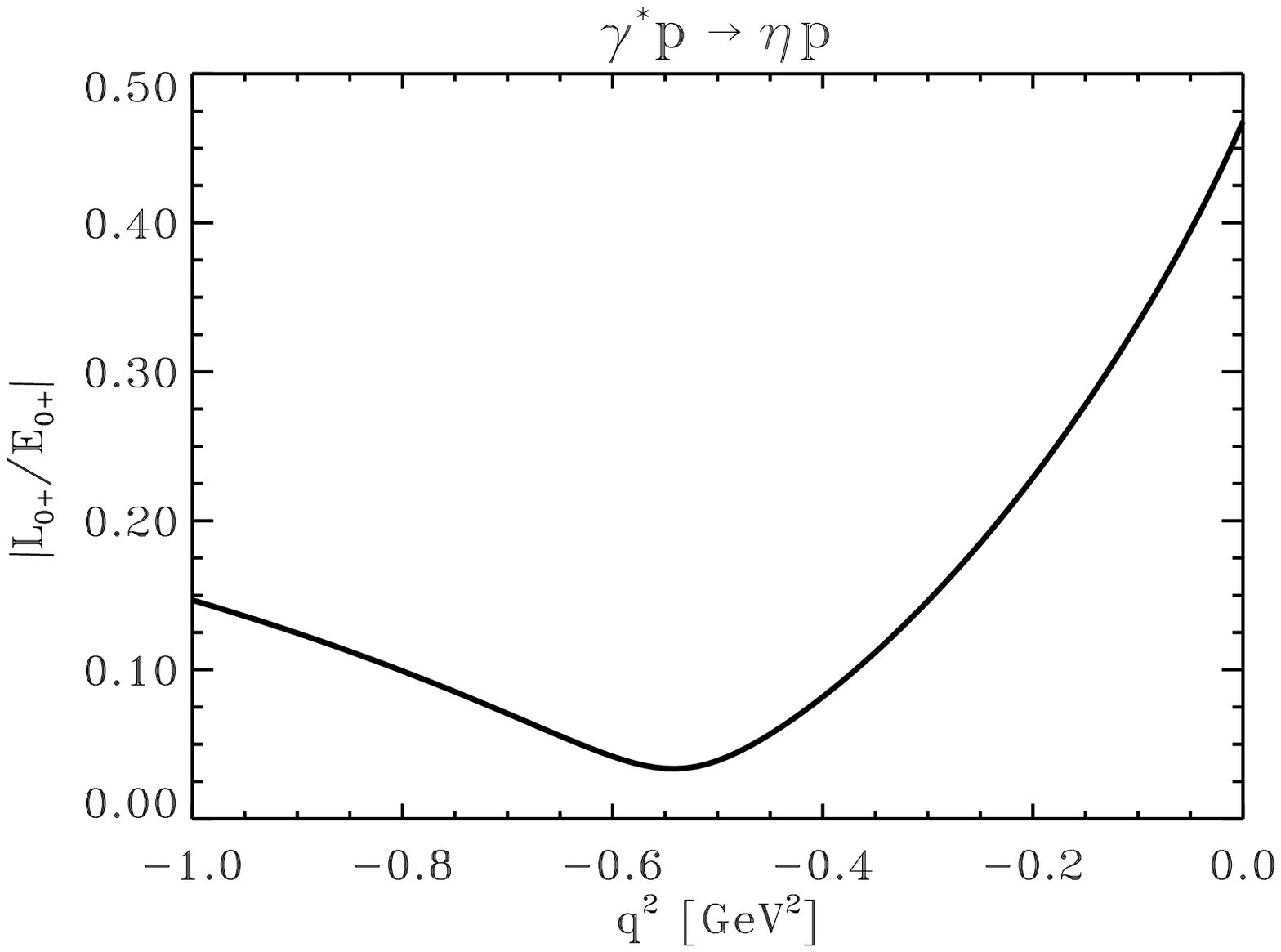}{16}
\vspace*{.5cm}
\centerline{\large Figure 5}

\newpage

\bild{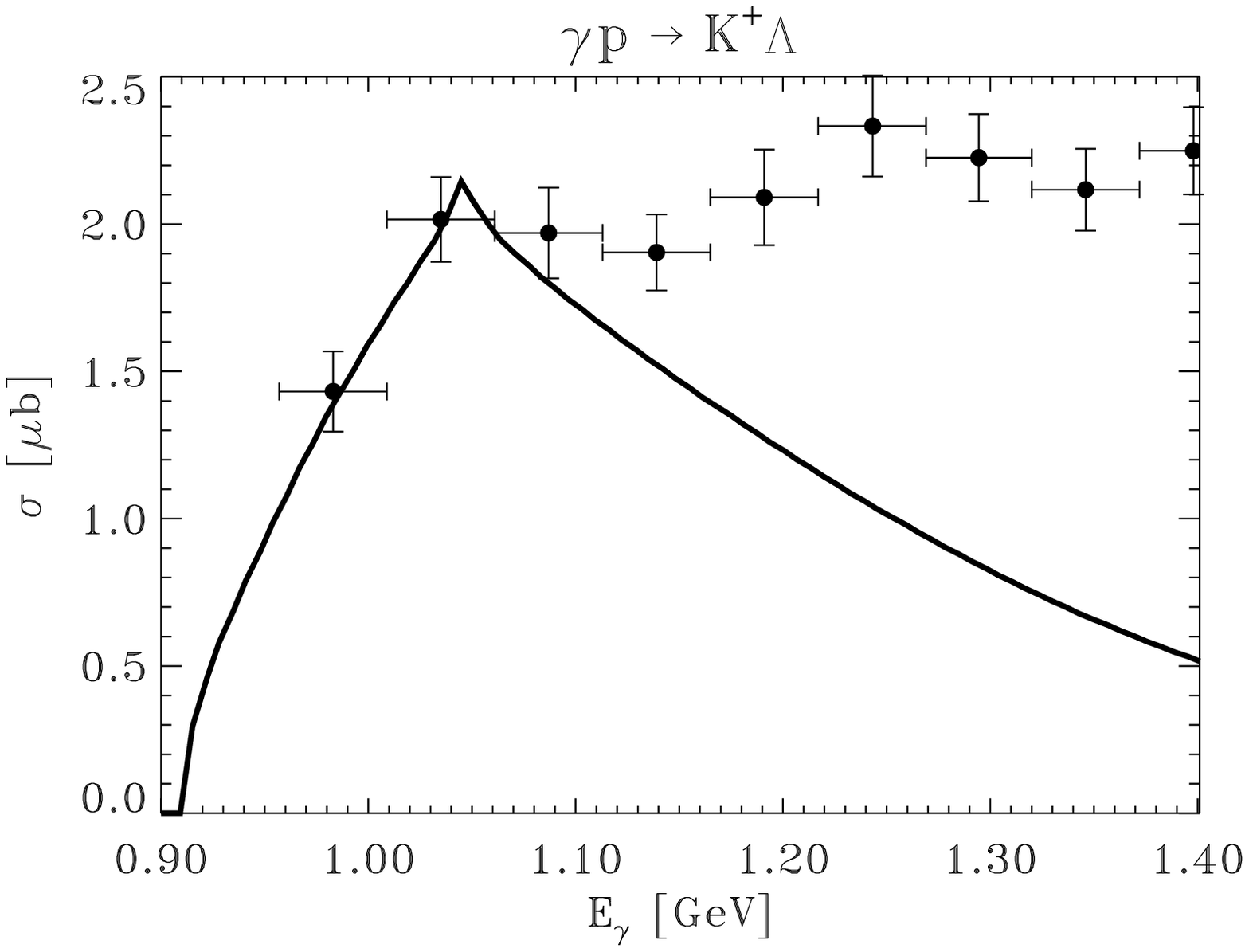}{10.5}
\bild{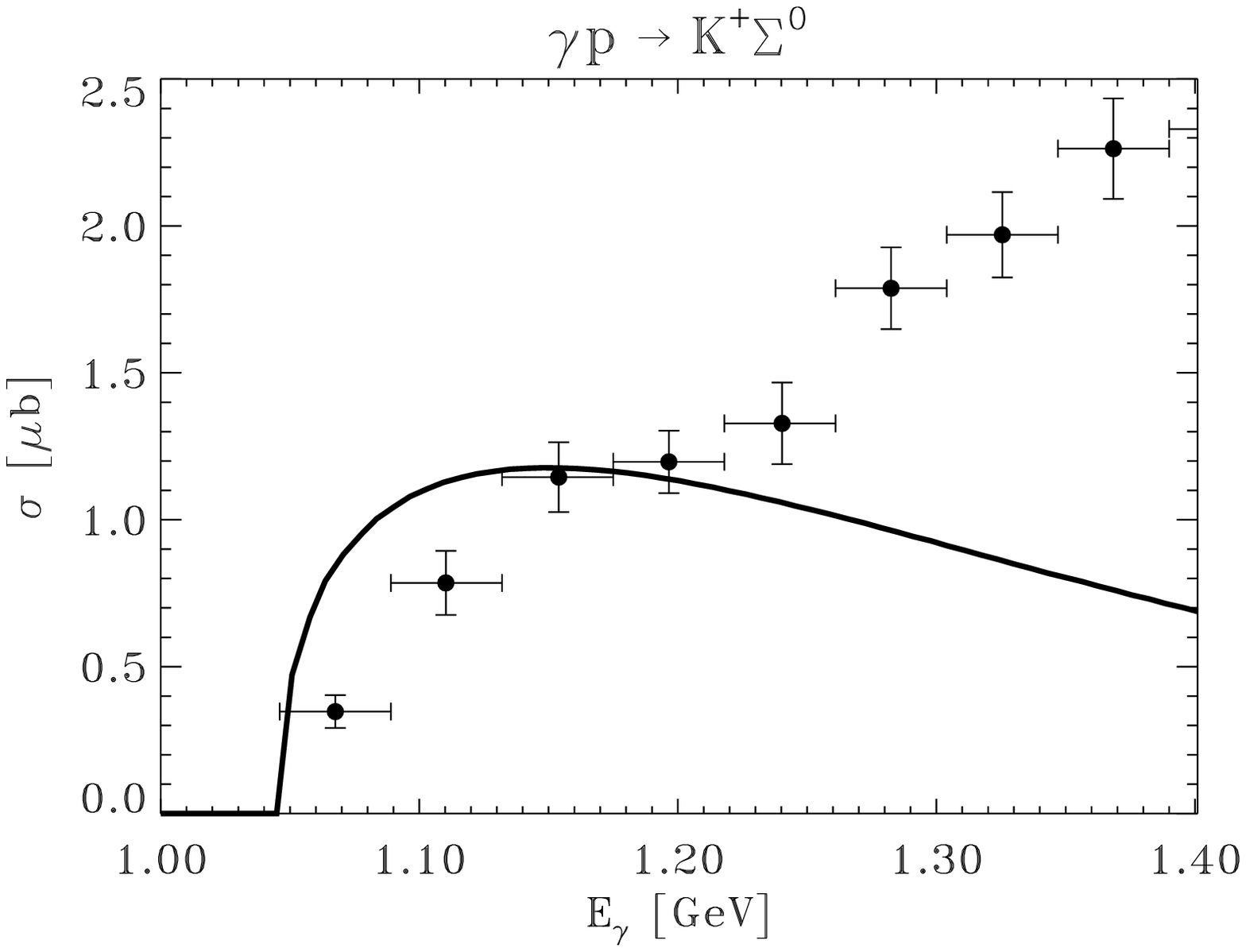}{10.5}
\bild{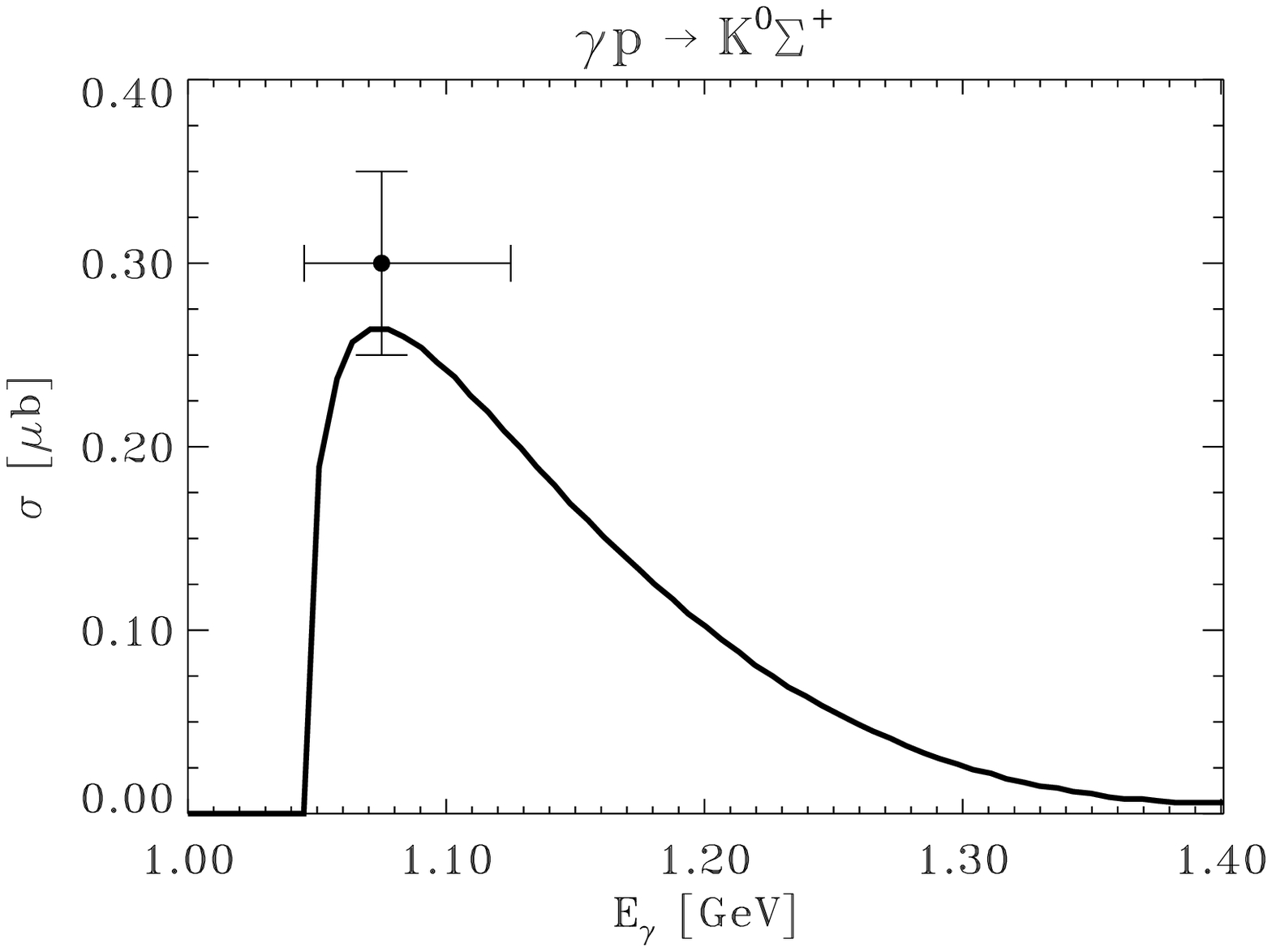}{10.5}

\centerline{\large Figure 6}

\newpage

\vspace*{2cm}
\bild{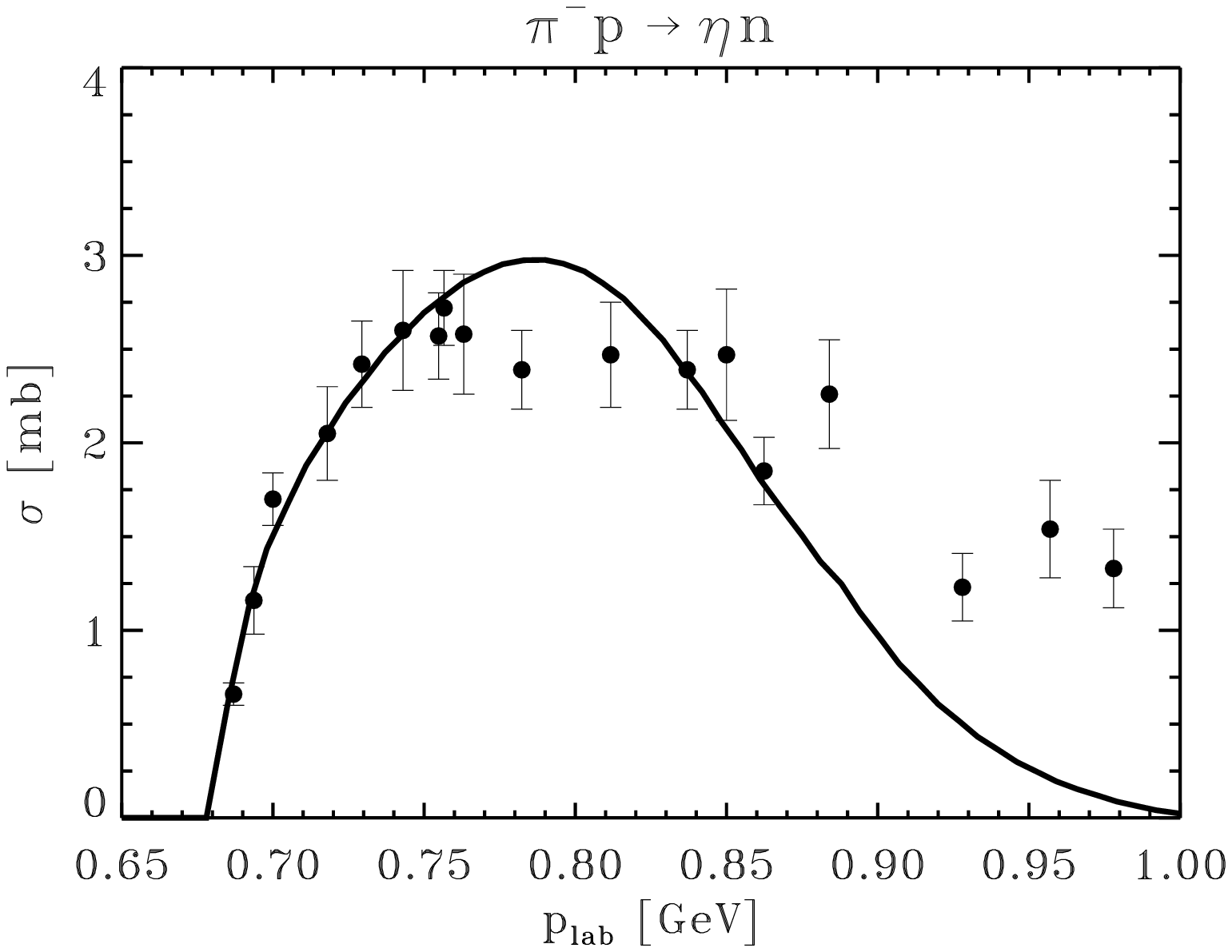}{7.5}
\begin{minipage}{7cm}
\bild{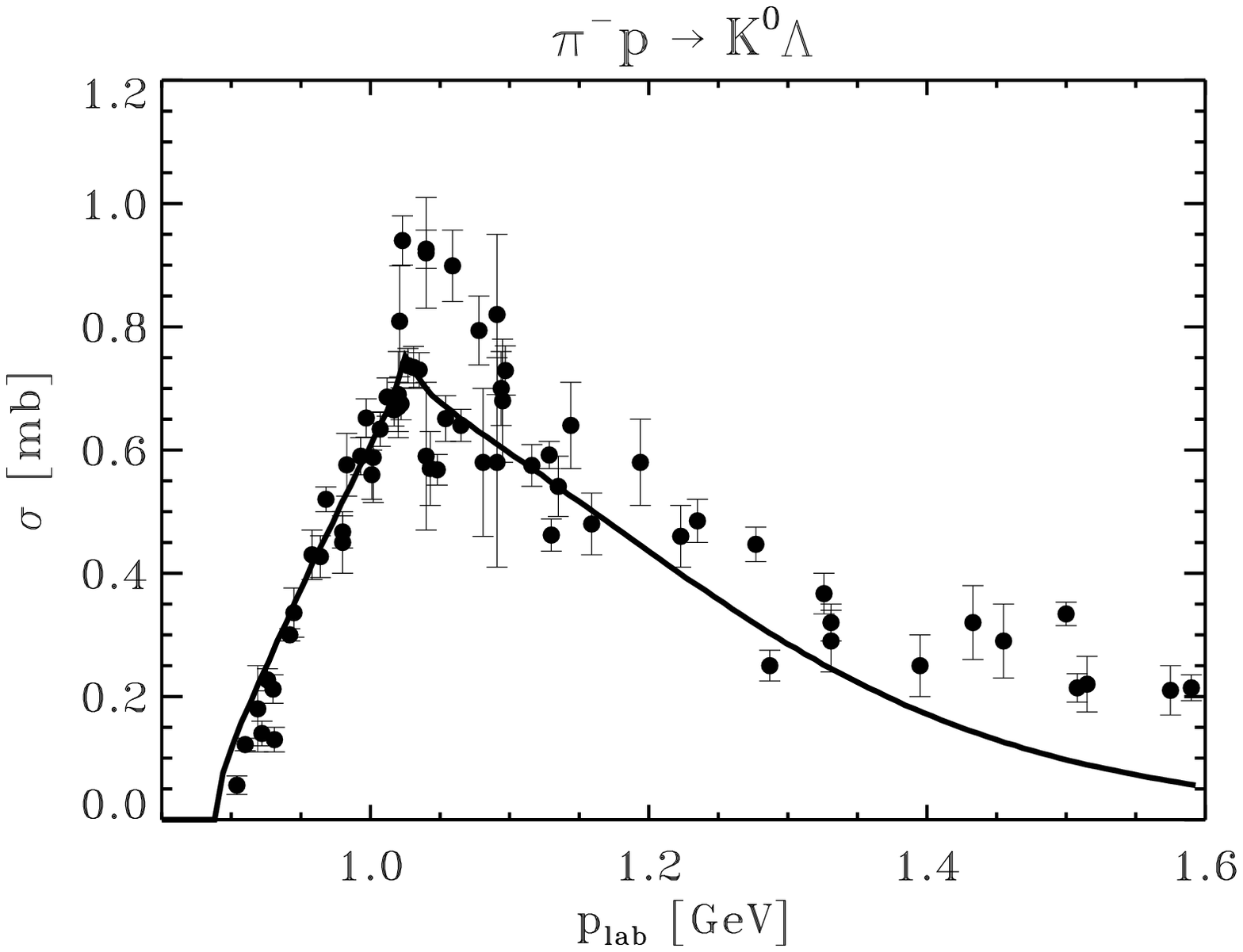}{7.5}
\bild{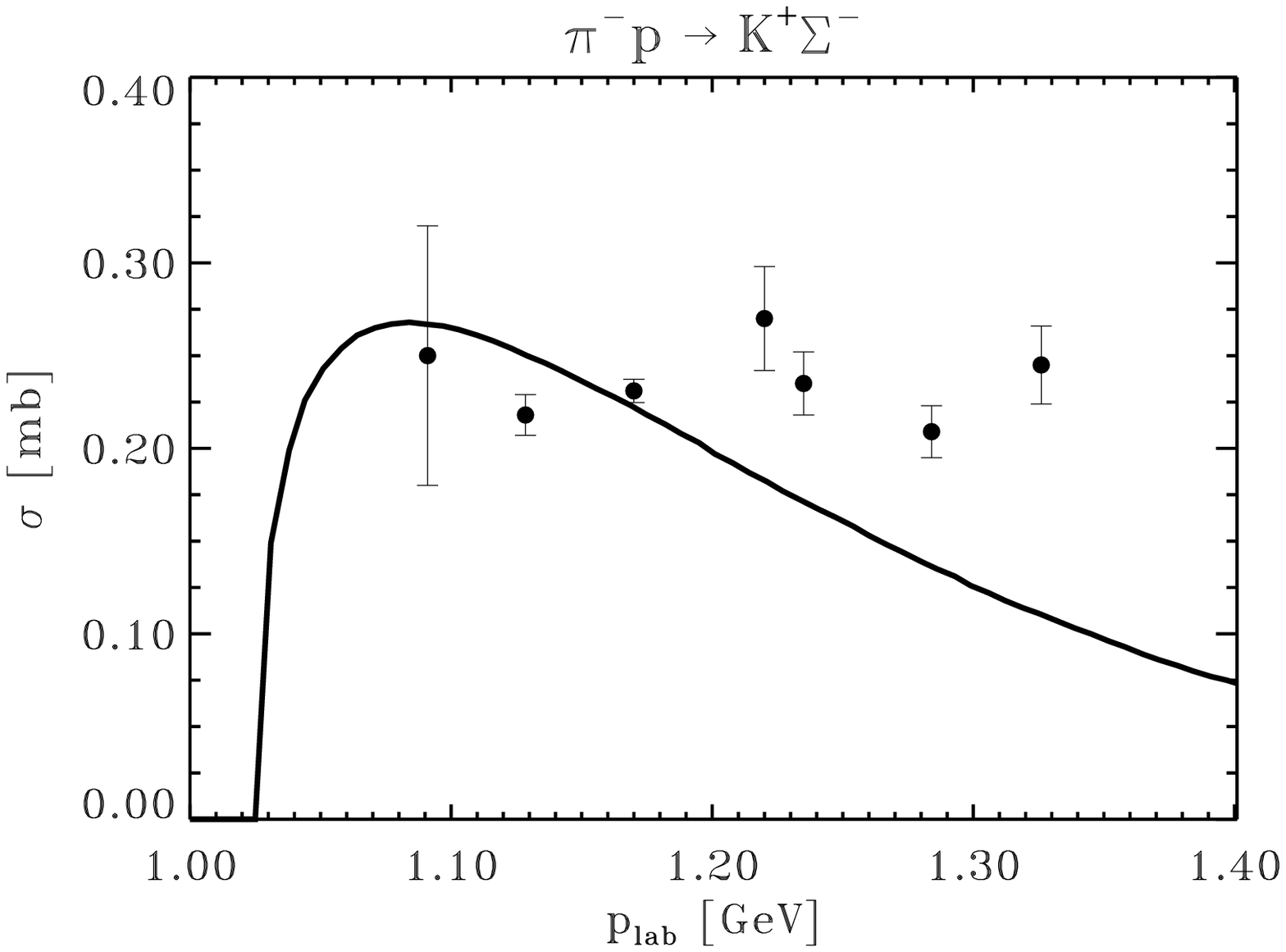}{7.5}
\end{minipage}
\begin{minipage}{7cm}
\bild{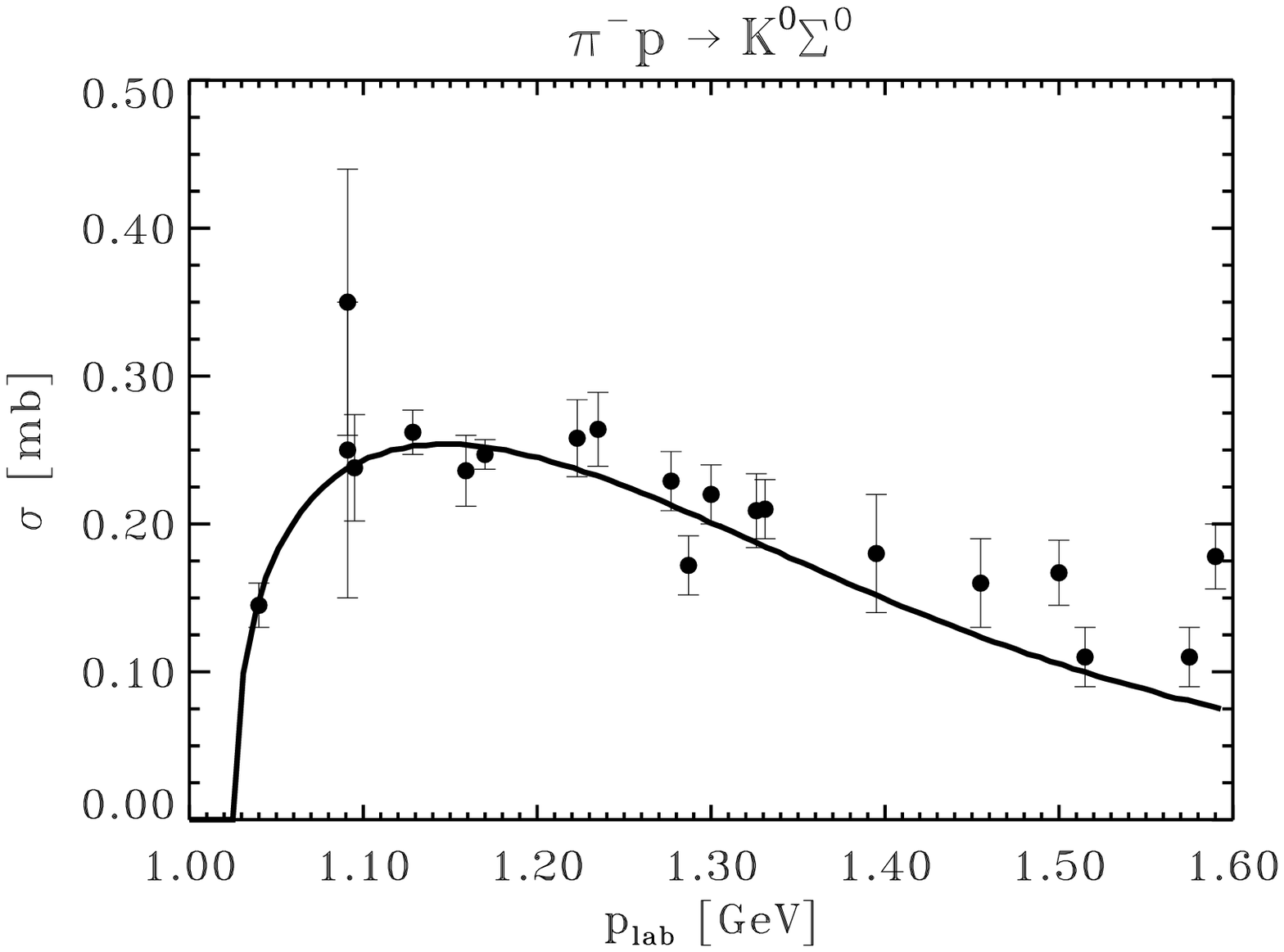}{7.5}
\bild{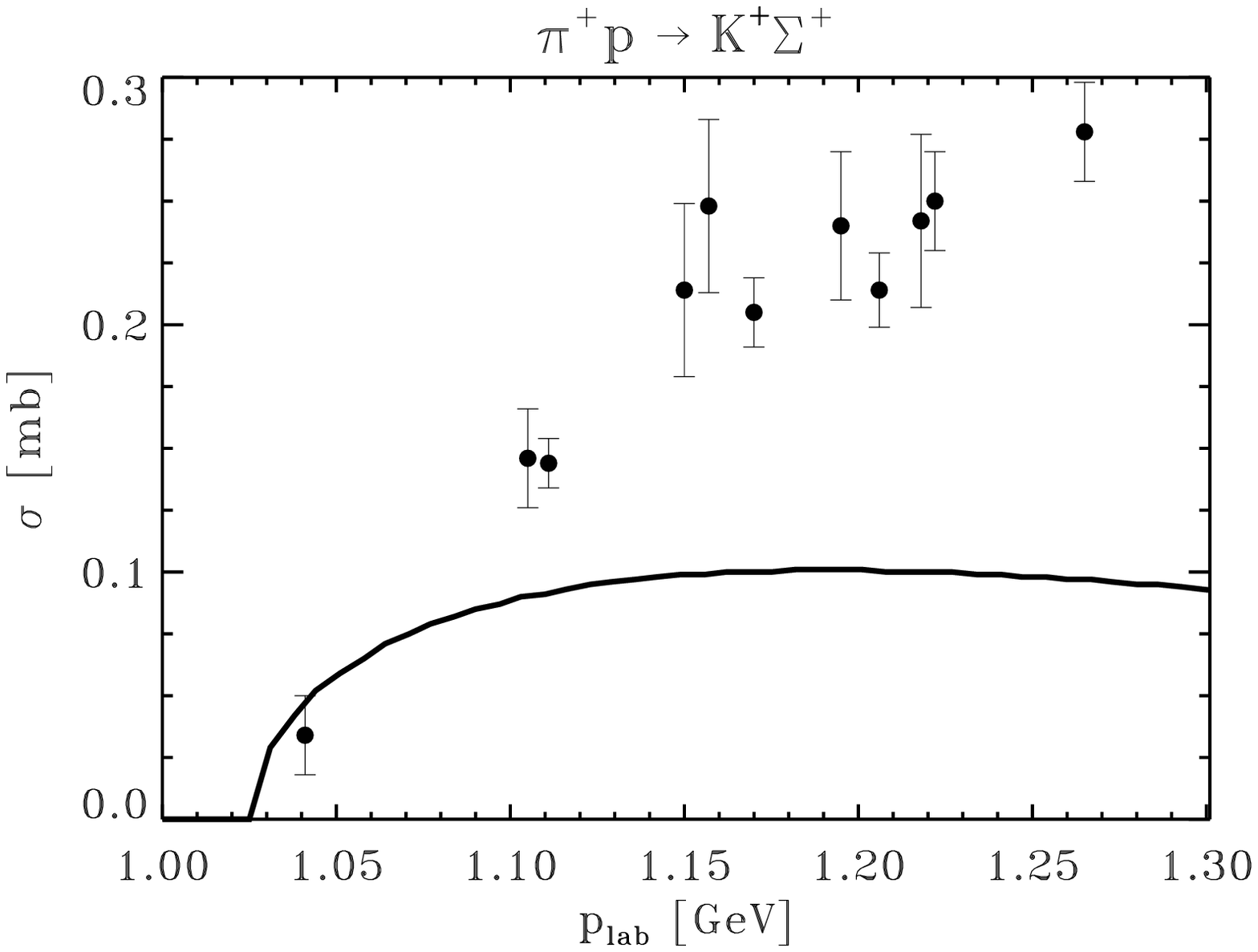}{7.5}
\end{minipage}

\vspace*{1cm}
\centerline{\large Figure 7}

\newpage

\bild{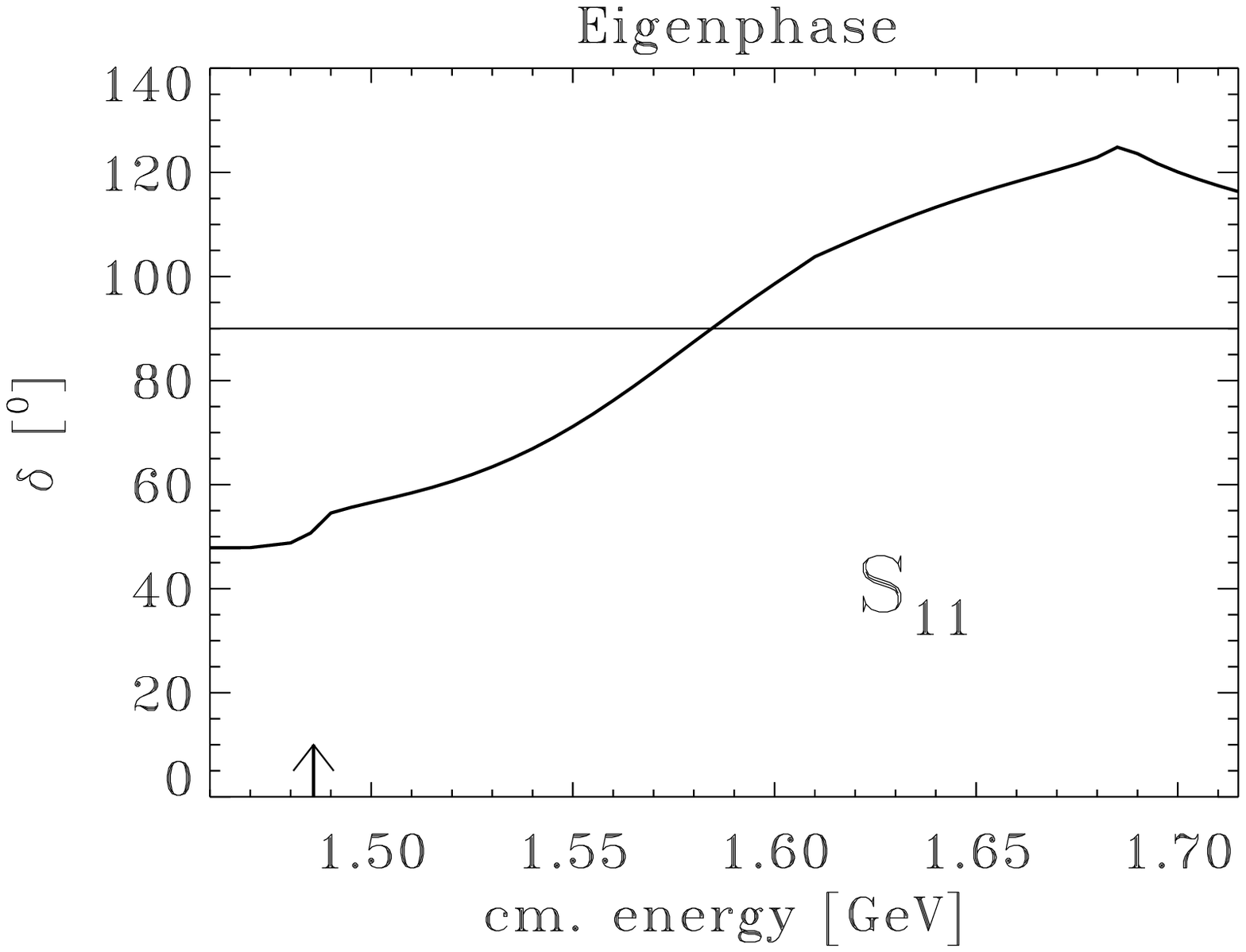}{16}
\vspace*{1cm}
\centerline{\large Figure 8}

\newpage

\bild{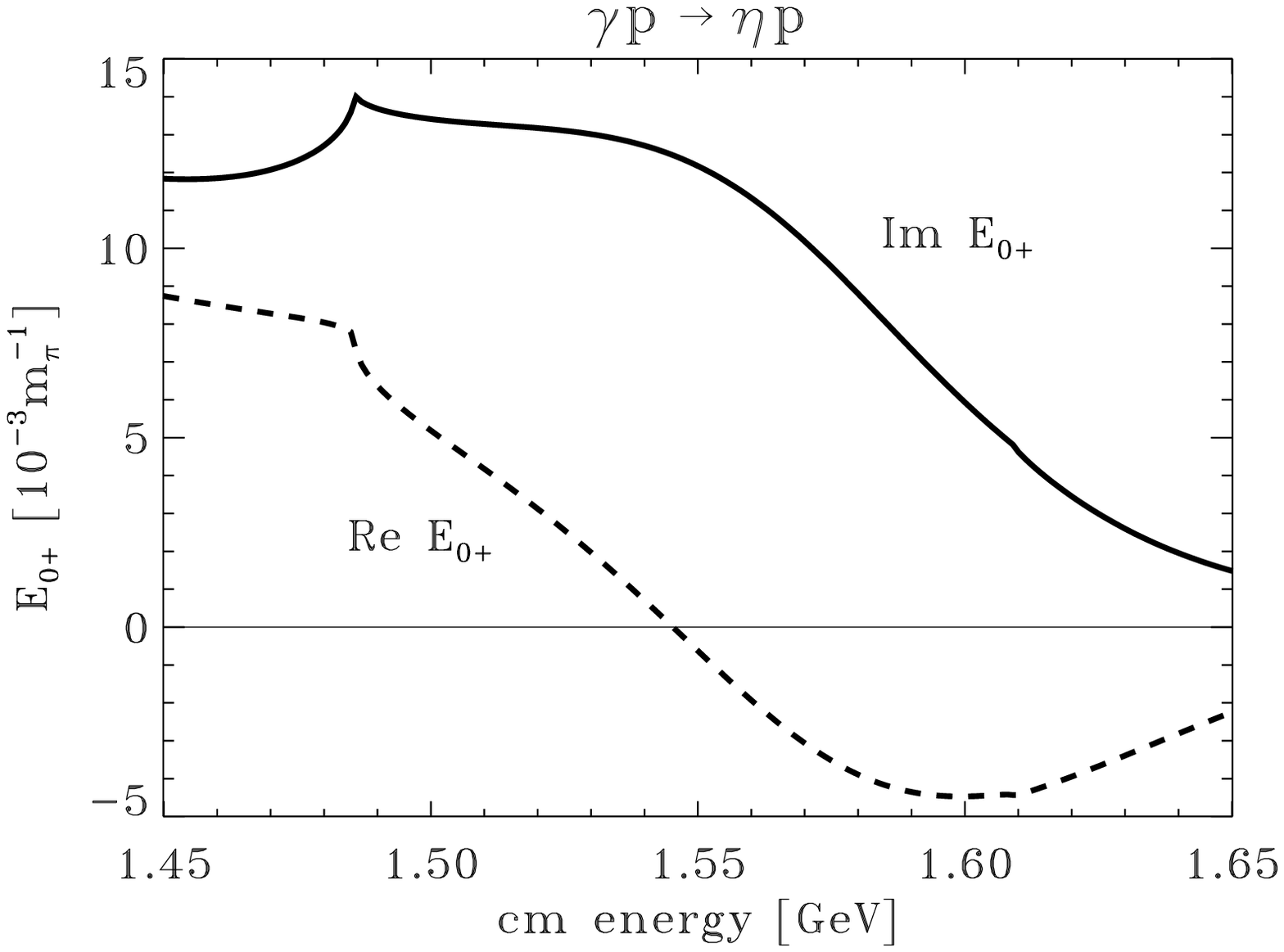}{16}
\vspace*{1cm}
\centerline{\large Figure 9}
\end{document}